\documentclass[a4paper,fleqn,usenatbib]{mnras}
\usepackage{amsfonts}
\usepackage[T1]{fontenc}
\usepackage{mathtools} 
\usepackage{graphicx}	
\usepackage{amsmath}	
\usepackage{amssymb}	
\usepackage{enumerate}
\def\shortauthor#1{\def\@shortauthor{#1}}

\newdimen\dummy
\dummy=\oddsidemargin
\newcommand{\vecv}{\mbox{\boldmath $v$} {}}
\newcommand{\vecx}{\mbox{\boldmath $x$} {}}
\newcommand{\vecz}{\mbox{\boldmath $z$} {}}
\newcommand{\vecF}{\mbox{\boldmath $F$} {}}
\newcommand{\vecbeta}{\mbox{\boldmath $\beta$} {}}
\newcommand{\nablavec}{\mbox{\boldmath $\nabla$} {}}

\input{tcilatex}
\begin{document}

\title[Mass-losing and mass-gaining perturbers]{Gaseous wakes and dynamical friction: mass-losing and mass-gaining perturbers}
\author[S\'anchez-Salcedo \& Chametla]{F. J. S\'anchez-Salcedo$%
^{1}$\thanks{%
E-mail:jsanchez@astro.unam.mx} and R. O. Chametla$^{2}$ \\
$^{1}$Instituto de Astronom\'{\i}a, Universidad Nacional Aut\'{o}noma de M%
\'{e}xico, Ciudad Universitaria, Apt.~Postal 70 264, \\
C.P. 04510, Mexico City, Mexico\ \\
$^{2}$Escuela Superior de F\'{\i}sica y Matem\'{a}ticas, Instituto Polit\'{e}%
cnico Nacional, UP Adolfo L\'opez Mateos, Mexico City, Mexico}

\date{Accepted xxxx Month xx. Received xxxx Month xx; in original form 2014
March 6}
\maketitle

\begin{abstract}
An extended gravitational object embedded in a parent system comprised of gas and
collisionless particles may undergo both dynamical friction (DF) and mass loss by tidal 
forces. If the object is compact enough, it can increase its mass through accretion of
material from the surrounding medium. 
We extend the classical linear analysis of DF on a constant-mass 
body in a gaseous medium to the case where its mass changes with time.
We show that the structure of the wake may differ significantly from the constant-mass case.
For instance, the front-back symmetry of density about subsonic constant-mass perturbers 
is broken down for variable-mass perturbers.
The density wake keeps a memory of the past mass history of the 
perturber. For dissolving perturbers, the density field is more dense than expected using the
instantaneous mass of the perturber in the classical formula. 
As a consequence, the instantaneous-mass approximation underestimates the drag force
for mass-losing perturbers and overestimates it for mass-gaining perturbers.
We present cases in which the percentage error in the drag force using the instantaneous-mass 
approximation is greater than $50\%$.

\end{abstract}


\section{INTRODUCTION}

A gravitational body moving through a field of light particles, or through 
a gaseous medium, experiences a drag force known as dynamical friction (DF), as a consequence
of the continuous gravitational deflection of field particles or fluid elements
\citep{cha43,bon44}.
DF may induce an exchange of angular momentum between the massive object and the background 
particles, which leads to an orbital spiraling of the object towards the centre of the potential.

DF in a gaseous medium may be relevant to describe the orbital evolution
of planetesimals or planets in eccentric or inclined orbits, when they
are still embedded in the gaseous protoplanetary disc 
\citep{mut11,rei12,can13,xia13,gri15}.
Common-envelope binary stars, or stars in gas-embedded star clusters
may also suffer orbital decay due to gaseous DF \citep{ric08,cha10,lei14}.
At subgalactic scales, gaseous DF is important during the early phases of
galaxy evolution \citep{ost99}.
It is also especially relevant in the central parts of gas-rich galaxies. 
Indeed, gaseous DF plays a key role in the orbital shrinking of supermassive
binary black holes in the central parts of gas-rich mergers of galaxies. 
Globular clusters and nuclear star clusters 
may also suffer significant migration due to DF with the gas component 
in gas-rich dwarf galaxies \citep[e.g.,][]{ant12,gui16}.

Most estimates of the DF force assume that the body is on
a straight-line trajectory at constant velocity through a homogeneous gaseous medium 
\citep[e.g.,][]{bon44,dok64,rud71,bis79,rep80,ost99,edg04,can13}
In recent years, the DF force has been studied adding other physical phenomena or 
adopting different assumptions.
\citet{kim07} computed the density wake and the DF on perturbers in 
circular orbits. \citet{san14} investigated the drag force
on a binary system and the torques on each component of the binary.
The DF force on a perturber travelling in
a magnetized gas has been studied in \citet{san12}
and \citet{sha12}.
\citet{nam10} relaxed the assumption that the
body moves at constant velocity and derived the DF on decelerating perturbers.
On the other hand, \citet{lee11} showed that, for subsonic perturbers, 
the drag force magnitude depends on whether the perturber can accrete mass or not.

In all the abovementioned studies, it is generally assumed that the perturber has constant
mass. However, in many scenarios, astrophysical bodies may lose or gain material and 
thereby its mass changes. For instance, planets embedded
in protoplanetary discs, may undergo rapid mass accretion at a rate that can
be understood in terms of accretion within the Bondi radius or within the
Hill radius, depending on its mass \citep[e.g.,][]{dan08}.
On the other hand, in star-forming molecular clouds, high-mass stars may form due to continued 
accretion of gas funneling to the centre of the cluster potential \citep{bon06}.
The mass function of star clusters may also indicate that clusters also grow by 
accretion \citep{kuz17}.

Extended objects, such as star clusters and satellite galaxies, can lose mass
due to stripping by the tidal field of the host galaxy or by ram pressure.
Especially dwarf galaxies may experience rapid mass loss when stellar feedback
blows out a large fraction of gas \citep[][]{gov10}.
As quantitatively described by \citet{zha04}, the strength of the tidal forces
on satellite galaxies and star clusters increase as they sink towards the
galactic centre due to DF. On some occasions, tidal forces may lead to a
complete disruption of the stellar cluster or satellite. According
to the $\Lambda$CDM models of \citet{fat18}, some
dwarf spheroidal galaxies in the haloes of the Milky Way and Andromeda
have been very heavily stripped. For instance, the fraction of mass lost
in Crater 2 and Andromeda XIX is about $99$ percent.

Since the strength of the DF force depends on the mass of the perturber 
(or satellite), the DF timescale, that is the time to reach the host's centre 
from a certain initial radius, depends not only on the initial mass of the 
perturber, but also on the mass loss rate \citep[e.g.,][]{col99,gan10}.
As a first approximation, one should use the instantaneous bound mass of the perturber 
to estimate the DF force. However, the mass in the tidal debris may also affect DF:
stripped material that remains in the vicinity of the perturber  
also contributes to the DF because they gravitationally interact with the
bound stars \citep{fuj06,fel07}.
This material also disperses background particles and enhances the amplitude of the wake 
behind the perturber. These tidal debris effects are important for stellar
systems undergoing tidal disruption \citep[e.g.,][]{fel07}.

In this work we study a different aspect of the DF force acting on a 
body of changing mass. Our analysis pivots on the fact that the wake excited 
in ambient medium keeps a memory of the history of the mass of the perturber. 
Since the drag force arises from the gravitational attraction between the body 
and its induced wake, the drag force should also reflect the history dependent
nature of the wake. Our aim is to estimate the DF on a variable-mass perturber as well as to
quantify how much it differs from the drag force derived using the instantaneous mass
of the perturber.

The paper is organized as follows. In Section \ref{sec:basics}, we describe the linear
hydrodynamical approach to derive the gaseous wake in the medium and
the DF force. In Section \ref{sec:wake}, we describe the structure of the wake induced by
a body with non-constant, continuously varying mass. Computations of the DF force are given
in Section \ref{sec:dynamical_friction}. Finally, we give a brief discussion and the conclusions in
Section \ref{sec:summary}.

\section{Gravitational wakes: basics and model}
\label{sec:basics}
\subsection{Variable-mass perturber}

We consider a gravitational body moving through a gaseous medium. The gas will respond to the 
gravitational pull created by this object. The perturber's 
gravitational potential, $\Phi_{p}(\vecx,t)$, satisfies the Poisson equation:
\begin{equation}
\nabla^{2}\Phi_{p} (\vecx,t)= 4\pi G \rho_{p}(\vecx,t),
\end{equation}
where $\rho_{p}(\vecx,t)$ is the density profile of the perturber.
The temporal evolution of $\rho_{p}(\vecx,t)$ depends 
on how the process of mass gain and/or mass stripping occurs. 
For simplicity, we assume that
the shape of the density profile of the perturber does not change with time,
so that the perturber's profile factorizes as 
\begin{equation}
\rho_{p}(\vecx,t)=\eta (t) \rho_{p,0}(\vecx-\vecx_{p}),
\label{eq:rhop}
\end{equation}
where $\vecx_{p}(t)$ is the position of the centre of mass of the perturber.
If $\eta$ is taken as a constant value, then we recover the constant-mass case.
The mass that remains bound to, or in the vicinity of the pertuber, is
\begin{equation}
M_{p}(t)= \eta (t) M_{0},
\label{eq:MpM0}
\end{equation}
where $M_{0}$ is a constant with dimensions of mass.
Thus, the mass of the perturber changes at a rate given by 
$\dot{M}_{p}=\dot{\eta} M_{0}$.

To quantify the DF force, we adopt a rather generic function 
for $\eta (t)$. 
We will assume that the perturber is formed at $t=-t_{0}$ (where
$t_{0}>0$), then it evolves at constant mass in the interval $-t_{0}<t<0$.
At $t=0$, it starts a phase of exponential mass loss or mass gain with a characteristic 
timescale $\tau$. 
Such a situation can be described by the function
\begin{equation}
\eta (t)=\left\{ 
\begin{tabular}{l}
$0$  \hskip 3.0cm at $t<-t_{0}$\\
$1$ \hskip 3.0cm  at $-t_{0}<t<0$ \\  
$1+ \lambda - \lambda \exp (-t/\tau)$ \ \ \  at $t>0$,
\end{tabular}
\right.  
\label{eq:piece_wise}
\end{equation}
with $\lambda\geq -1$.
A mass model is specified by four parameters: $M_{0}$, $t_{0}$, $\tau$ and $\lambda$.
Here $\lambda$ is the fraction of mass that the perturber has lost or gained
since $t=0$ to $t\rightarrow \infty$. More specifically, the mass of the perturber
is $M_{0}$ at $t=0$, and it is $(1+\lambda)M_{0}$ at $t\rightarrow \infty$. Values in the 
range $-1<\lambda<0$ correspond to mass loss. For models with $\lambda>0$, the perturber
enhances its mass. In a more compact notation, $\eta$ can be written as
\begin{equation}
\eta (t)= \Theta (t+t_{0}) +\lambda [1-\exp \left(-t/\tau\right)] \Theta (t),
\label{eq:etaplus}
\end{equation}
where $\Theta(t)$ is the Heaviside function.
If we adopt $\lambda=0$ or $\tau\rightarrow \infty$, it describes a case where the perturber
is turned on at $t=-t_{0}$ and remains with constant mass $M_{0}$
at later times. This case was studied by \citet{ost99}.

\begin{figure*}
\centering
 \includegraphics[scale=1.7,height=95mm,width=0.75\textwidth]{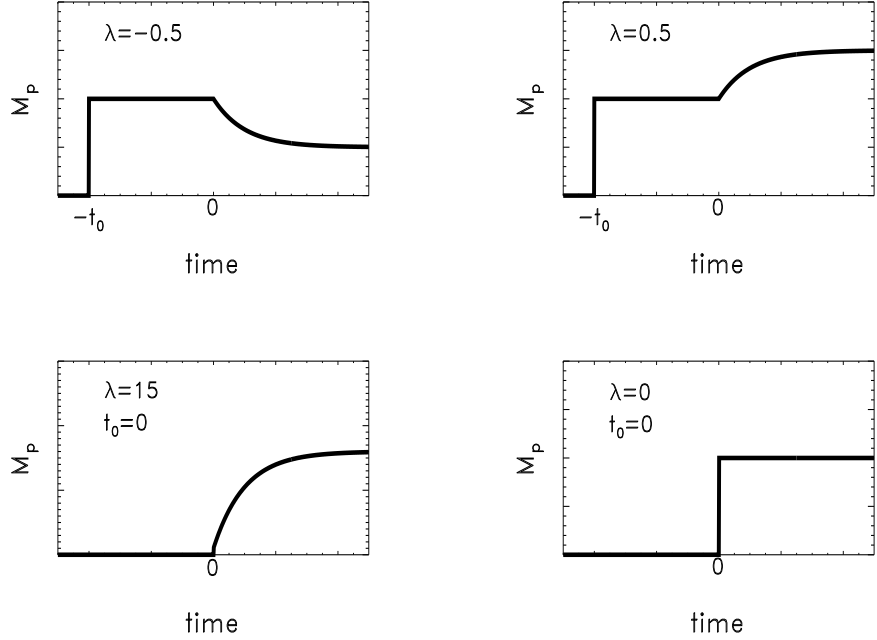}
 \caption{Some representative cases of the temporal evolution of the mass of the perturber.}     
 \label{fig:representative}
\end{figure*}

Figure \ref{fig:representative} shows a sketch of four representative models.
The meaning of $t_{0}$
depends on the setting and configuration of the system under study.
For instance, if we are interested in studying the DF on
a forming object as that depicted in the third panel of Figure \ref{fig:representative},
we may choose $t_{0}=0$. In other astrophysical settings, 
$t_{0}$ may represent the time elapsed since the body has entered 
into the gas medium until it experiences mass loss. In fact,
tidal forces do not lead to significant mass stripping if the size of the body
is smaller than the tidal radius. Once the tidal radius has been filled due to
tidal heating, the satellite may suffer important tidal stripping.

A period of almost constant mass and a subsequent episode of mass loss, as the scenario
shown in the first panel of Figure \ref{fig:representative}, are easily recognizable in simulations
of tidal evolution of satellite systems; tidal effects are more intense
at pericentre passages, where episodes of severe mass loss occur  \citep[e.g.,][]{pen10}.
We will return to the discussion of the physical meaning of $t_{0}$ in \S \ref{sec:summary}.

Although our choice of $\eta (t)$ gives enough leeway to explore
the significance of different parameters, other variants may be also
relevant. For instance, collisional $N$-body calculations of isolated
star clusters show an almost linear decrease of mass with time \citep{bau01}. 
In Appendix \ref{sec:appD}, more generic choices for the mass evolution of the perturber
are considered.

\subsection{Linear equations and formal solution}
\label{sec:formal_solution}
To derive the density structure of the wake induced by the perturber, 
we will follow the same approach as described in \citet{ost99}.
The unperturbed gaseous medium is homogenous and infinite, with density
$\rho_{\infty}$ and sound speed $c_{s}$. The gas is initially at rest.
Far enough from the perturber, the disturbances in gas density
and velocity are always linear. Close to the perturber, the perturbations might be
also linear if the characteristic size of the perturber is much larger than the accretion
radius defined as $GM_{p}/(V^{2}+c_{s}^{2})$, where $V$ is the velocity of 
the perturber (see \S \ref{sec:wake}).

We define the perturbed gas density as $\alpha \equiv (\rho-\rho_{\infty})/\rho_{\infty}$ and
the perturbed gas velocity as $\vecbeta=\vecv/c_{s}$. Provided that $\alpha\ll 1$ and $\beta\ll 1$,
the linearized Euler equations describe the evolution of the system.
In terms of $\alpha$ and $\vecbeta$ they can be written as
\begin{equation}
\frac{1}{c_{s}}\frac{\partial \alpha}{\partial t}+\nablavec \cdot\vecbeta=0,
\end{equation}
\begin{equation}
\frac{1}{c_{s}}\frac{\partial \vecbeta}{\partial t}+\nablavec \alpha=-\frac{1}{c_{s}}
\nablavec \Phi_{p}.
\end{equation}
By combining these two equations, it is simple to show that $\alpha (\vecx,t)$ satisfies 
the following inhomogeneous wave equation 
\begin{equation}
\nabla ^{2}\alpha -\frac{1}{c_{s}^{2}}\frac{\partial ^{2}\alpha }{\partial
t^{2}}=-\frac{4\pi G}{c_{s}^{2}} \rho_{p}\left(\vecx,t\right).
\end{equation}
The solution to this equation is
\begin{equation}
\alpha \left(\vecx,t\right) =\frac{G}{c_{s}^{2}}\iint d^{3}x^{\prime
}dt^{\prime }\frac{\delta \left[ t^{\prime }-\left( t-\left\vert 
\vecx-\vecx^{\prime}\right\vert /c_{s}\right) 
\right] \rho_{p}\left(\vecx^{\prime },t^{\prime }\right) }{\left\vert 
\vecx-\vecx^{\prime }\right\vert }.
\label{eq:linearalpha}
\end{equation}
Once $\rho_{p}(\vecx,t)$ is specified, we may compute $\alpha(\vecx,t)$.
In our case and according to Equations (\ref{eq:rhop}) and (\ref{eq:etaplus}), $\rho_{p}$ 
is the sum of three terms
\begin{equation}
\rho_{p}(\vecx,t)= \sum_{i=1}^{3} \rho_{p,i},
\end{equation}
with
\begin{equation}
\rho_{p,1}(\vecx,t)=\Theta(t+t_{0})\rho_{p,0}(\vecx-\vecx_{p}),
\end{equation}
\begin{equation}
\rho_{p,2}(\vecx,t)=\lambda \Theta(t)\rho_{p,0}(\vecx-\vecx_{p}),
\end{equation}
\begin{equation}
\rho_{p,3}(\vecx,t)=-\lambda\Theta(t)\exp(-t/\tau) \rho_{p,0}(\vecx-\vecx_{p}).
\end{equation}
Since Equation (\ref{eq:linearalpha}) is linear, the perturbed density $\alpha(\vecx,t)$
is a superposition of each individual solution. More specifically, if $\alpha_{i}$ is the 
perturbed density created by a perturber with a density profile $\rho_{p,i}$, then $\alpha=\sum \alpha_{i}$. 
For shortness, we will refer to the wake associated with the mass density term
$\rho_{p,1}$ as the wake 1, and so on.

Once $\alpha$ is computed, the gravitational drag felt by the perturber is given by
\begin{equation}
\vecF_{\rm DF}= \rho_{\infty}\int \alpha \nablavec \Phi_{p} \, d^{3}\vecx.
\label{eq:DF_original}
\end{equation}
As usual, the integral is performed over all volume excluding a sphere of radius $r_{\rm min}$ around 
the perturber, where $r_{\rm min}$ is the distance from
the perturber at which the linear approximation breaks down.

\subsection{Particular case: Constant-mass perturber}
\citet{dok64}, \citet{rud71} and \citet{rep80}
calculated, in linear theory, the density structure of the stationary wake induced by a gravitational object 
of constant mass $M_{0}$, moving in a rectilinear orbit with velocity $V\hat{\vecz}$. 
The stationary case corresponds to $\eta=1$ in our notation.
The gas response depends on the Mach number of the perturber, 
defined as $\mu\equiv V/c_{s}$. In the stationary wake,
subsonic perturbers generate density distributions whose isodensity contours are ellipsoids centred
on the perturber. Therefore, the net DF force on subsonic perturbers is zero in the steady-state flow.
For supersonic bodies, the magnitude of the DF force is given by
\begin{equation}
F_{\rm DF}=\frac{4\pi\rho_{\infty}(GM_{0})^{2}}{V^{2}} \ln \Lambda,
\label{eq:general_form}
\end{equation}
where $\ln\Lambda\equiv \ln r_{\rm max}/r_{\rm min}$ is the Coulomb logarithm.
Here $r_{\rm min}$ and $r_{\rm max}$ are the minimum and maximum radii of the effective
gravitational interaction of a perturber with a gas.

\citet{ost99} performed the time-dependent analysis of the wake produced by
a perturber that is turned on at $t=0$ \citep[see also][]{jus90}.
She found that the perturbed density in the wake is
\begin{equation}
\alpha(\vecx,t)=\frac{GM_{0}}{c_{s}^{2}D}\xi_{0},
\end{equation}
where $D\equiv [s^{2}+(1-\mu^{2})R^{2}]^{1/2}$, with $R=(x^{2}+y^{2})^{1/2}$ the cylindrical radius,
$s\equiv z-Vt$ and
\begin{equation}
\xi_{0}=\left\{ 
\begin{tabular}{l}
1 \ \ \ \ \ \  if $R^{2}+z^{2}<(c_{s}t)^{2}$,\\ 
2 \ \ \ \ \ \  if $\mu>1$, $R^{2}+z^{2}>(c_{s}t)^{2}$,\\
  \ \ \ \ \ \ \ \ \ \ $s/R<-(\mu^{2}-1)^{1/2}$, and $z>c_{s}t/\mu$,\\
$0$ \ \ \ \ \ \   otherwise.
\end{tabular}
\right.
\end{equation}

\citet{ost99} noticed that the finite-time perturbation is more appealing because it 
captures more relevant physics than the stationary approach. 
For instance, she showed that the DF force is nonzero even for subsonic 
perturbers. The magnitude of the force for subsonic perturbers is given by 
Equation (\ref{eq:general_form}) with
\begin{equation}
\ln\Lambda=\frac{1}{2}\ln \left(\frac{1+\mu}{1-\mu}\right)
-\mu,
\label{eq:ostriker1}
\end{equation}
provided that $t>r_{\rm min}/(c_{\infty}-V)$.
In addition, the time-dependent analysis allows to remove the ambiguity in the definition
of $r_{\rm max}$ and permits to find the temporal behaviour of the Coulomb logarithm.
For supersonic perturbers and at $t>r_{\rm min}/(V-c_{\infty})$, she found
\begin{equation}
\ln\Lambda=
\ln\left(\frac{Vt}{r_{\rm min}}\right) +\frac{1}{2}\ln \left(1-\mu^{-2}\right).
\label{eq:ostriker2}
\end{equation}

\begin{figure*}
 \vbox{\vspace{-10.7em}\includegraphics[width=0.9\textwidth,height=160mm]{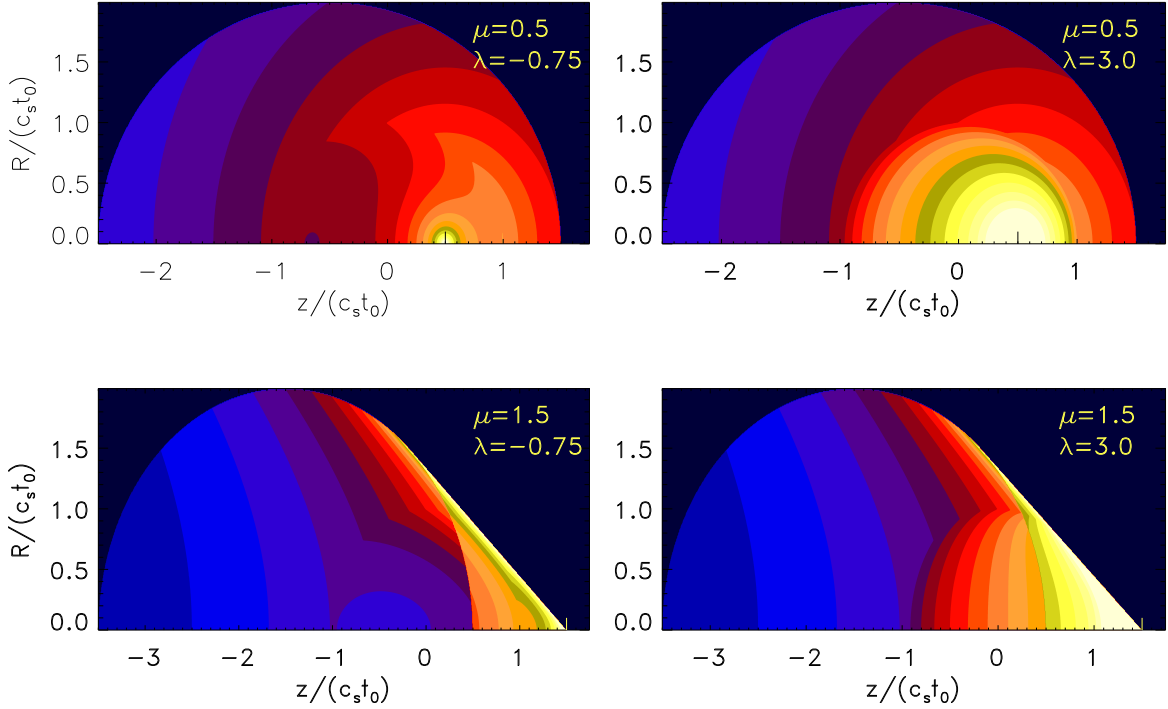}\vspace{-4.0em}}
 \caption{Isodensity contours at $t=t_{0}$ in the plane $(z,R)$. In all the cases, $\tau=t_{0}/2$. The long bar in
the $z$-axis indicates the position of the pertuber.}     
 \label{fig:map_density}
\end{figure*}

\section{The structure of the wake}
\label{sec:wake}
In the following, we present the density structure of the wake induced by a perturber whose mass 
varies over time according to Equations (\ref{eq:rhop}) and (\ref{eq:etaplus}).
We assume that the perturber moves at constant velocity
in a rectilinear orbit along the $z$-axis. The position of the centre of mass
of the body is $\vecx_{p}(t)=(0,0,Vt)$ and its density can be written as
\begin{equation}
\rho_{p}(\vecx,t)=\eta(t)\rho_{p,0}(x,y,z-Vt).
\end{equation}

In Appendix \ref{sec:appA} and \ref{sec:appB}, we evaluate the integral given in 
Equation (\ref{eq:linearalpha}) to obtain the analytical expressions for the three 
components of the wake ($\alpha_{1}$,
$\alpha_{2}$ and $\alpha_{3}$) excited by a point-mass perturber 
$\rho_{p,0}=M_{0}\delta(x)\delta(y)\delta(z-Vt)$.
The reliability of our analyical
derivation of $\alpha$ has been proven by comparing with the results of
hydrodynamical simulations (see Appendix \ref{sec:appC}).

The wakes $\alpha_{1}$ and $\alpha_{2}$ are
Ostriker type wakes; $\alpha_{1}$ corresponds to the wake created by a 
perturber of mass $M_{0}$ formed at $t=-t_{0}$, and $\alpha_{2}$
is the wake created by a fictitious perturber of mass $\lambda M_{0}$ formed at $t=0$ (wake 2).
The solution for $\alpha$ is:
\begin{equation}
\alpha(\vecx,t)=\frac{GM_{0}}{c_{s}^{2}D}[\xi_{1}+\lambda\xi_{2}-\lambda\exp(-t/\tau)\xi_{3}].
\label{eq:alpha_start}
\end{equation}
Since the analysis is linear, the above expression for $\alpha$ is only valid at those values of 
$D$ for which $\alpha\ll 1$.
The values for $\xi_{1}(\vecx,t)$ and $\xi_{2}(\vecx,t)$ are given 
in Appendix \ref{sec:appA}, whereas $\xi_{3}(\vecx,t)$ can be found in Equation (\ref{eq:xi3}).
The functions $\xi_{1}$, $\xi_{2}$ and $\xi_{3}$ depend on the position and time $(R,z,t)$, as
well as on the parameters $t_{0}$, $\mu$ and $\tau$. 
As shown in the Appendices \ref{sec:appA} and \ref{sec:appB}, the functions $\xi_{1}$, $\xi_{2}$
and $\xi_{3}\exp(-t/\tau)$, take values between $0$ and $2$.

In a scenario where the perturber dissolves, it is natural to assume
that the perturber is extended. If the characteristic physical size of the perturber is larger than 
$\sim GM_{p}/(c_{s}^{2}+V^{2})$,
the response of the gas is linear at any position in space \citep[e.g.,][]{ber13}.
In that case, the perturbed density induced by a softened perturber, $\alpha_{\rm soft}$, can be 
calculated using the convolution theorem as
\begin{equation}
\alpha_{\rm soft}(\vecx,t)=\frac{1}{M_{0}} \int  \alpha(\vecx-\vecx',t) \rho_{p,0}(\vecx',t) d^{3}\vecx',
\label{eq:softened_alpha}
\end{equation}
where $\alpha(\vecx,t)$ is given in Equation (\ref{eq:alpha_start}).
Nevertheless, the perturbed density and velocity in the far field (at distances much larger than the size of the
perturber) are essentially the same for extended and point-mass perturbers.

\begin{figure*}
\centering
 \includegraphics[width=0.9\textwidth]{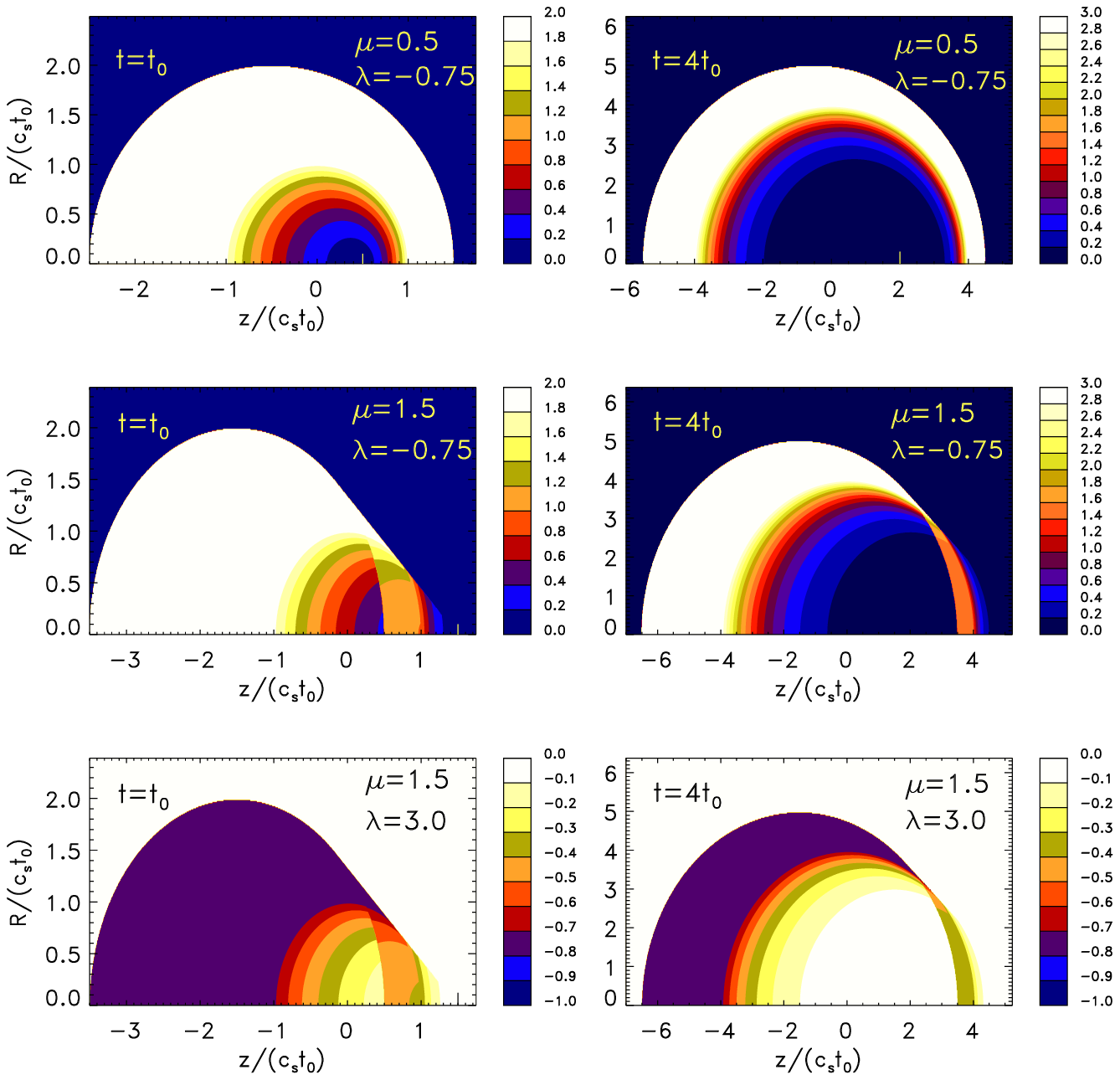}
 \caption{Isocontours of ${\mathcal{T}}$ for different combinations
of $\mu$ and $\lambda$, as indicated in each panel. In all cases, $\tau=t_{0}/2$. The long bar in
the $z$-axis indicates the position of the pertuber. In the panels corresponding to $t=4t_{0}$ and 
$\mu=1.5$, the perturber is located at $z=6$ (in units of $c_{s}t_{0}$) and hence it is outside
the range of the figure.}     
 \label{fig:map_sT}
\end{figure*}

Figure \ref{fig:map_density} shows colour maps of the perturbed density $\alpha$ at
time $t=t_{0}$, for different combinations
of $\mu$ and $\lambda$. In all the four cases, $\tau=t_{0}/2$. Hence the
current mass of the perturber is $0.35M_{0}$ and $3.59M_{0}$ for $\lambda=-0.75$ and $\lambda=3$,
respectively.  We see that the wake structure is
more complex than in the standard case $\lambda=0$. As said in \S 2.3,
the isodensity contours for subsonic constant-mass perturbers are ellipsoids.
However, as the model with $\lambda=3$ and $\mu=0.5$ illustrates,
the contours are not longer ellipsoids; some isodensity contours are flattened
along the line of motion of the perturber but others do not. Interestingly,
the backward-forward symmetry
of the wake excited by constant-mass subsonic
perturbers, is broken down for varying-mass perturbers (see also the map for
the model with $\lambda=-0.75$ and $\mu=0.5$). On the other hand, for supersonic
perturbers ($\mu>1$), the density wake is confined to the Mach cone and
sonic sphere, as occurs in the constant-mass case, but the isodensity
contours are not longer hyperbolae.

In order to quantify the imprint of time-dependent mass on the structure of the wake, 
we compare $\alpha(\vecx,t)$ with $\alpha_{\rm cst}(\vecx,t)$, defined as the perturbed 
density derived  in the constant-mass approximation. More specifically, $\alpha_{\rm cst}(\vecx,t)$ 
is the perturbed density simply adapting Ostriker's formula for a body created at $t=-t_{0}$, and taking 
the mass of the perturber as the instantaneous mass $M_{p}(t)$:
\begin{equation}
\alpha_{\rm cst}(\vecx,t)=\frac{GM_{p}(t)}{c_{s}^{2}D}\xi_{1}=\frac{G\eta(t)M_{0}}{c_{s}^{2}D}\xi_{1}.
\end{equation}
Recall that the wake factor $\xi_{1}$ is analogous to $\xi_{0}$, but for a body formed at $t=-t_{0}$.

At $t<0$, it holds that $\alpha=\alpha_{\rm cst}$. However,
at $t>0$, the difference between the exact value for the perturbed density $\alpha$, and 
the value derived in the instantaneous approximation is
\begin{equation}
\alpha-\alpha_{\rm cst}=-\frac{\lambda GM_{0}}{c_{s}^{2}D} [\xi_{1}-\xi_{2}+\exp(-t/\tau)(\xi_{3}-\xi_{1})].
\label{eq:diff_alpha1}
\end{equation}
Therefore, the fractional change of the density relative to $\alpha_{\rm cst}$ is
\begin{equation}
{\mathcal{T}}\equiv \frac{\alpha-\alpha_{\rm cst}}{\alpha_{\rm cst}}=-\frac{\lambda}{\eta\xi_{1}}
[\xi_{1}-\xi_{2}+(\xi_{3}-\xi_{1})\exp(-t/\tau)].
\label{eq:definition_sT}
\end{equation}
In regions where ${\mathcal{T}}$ is different from zero, the wake keeps a memory of the past
mass of the perturber.

Figure \ref{fig:map_sT} shows ${\mathcal{T}}$ for different combinations of parameters. 
It is a generic result that when the perturber dissolves (i.e. $\lambda<0$), ${\mathcal{T}}\geq 0$ at any point
in space because $\alpha > \alpha_{\rm cst}$; the vake is more dense
than predicted in the instantaneous approximation because the perturber was
more massive in the past. Conversely, when the perturber gains mass ($\lambda>0$),
${\mathcal{T}}\leq 0$ everywhere. In this case, the wake is less dense than $\alpha_{\rm cst}$.

For subsonic perturbers, $|{\mathcal{T}}|$ is larger in the outer parts of the wake than in the inner
parts. In fact, $|{\mathcal{T}}|$ decreases close to the body (see Fig. \ref{fig:map_sT}). 
For supersonic perturbers, $|{\mathcal{T}}|$ is also larger
in the outer parts of the wake that are far away from the body than in the near field region, but it does not 
drop monotonically when we approach from the outer wake towards the 
perturber; ${\mathcal{T}}$ exhibits a jump when we cross the sonic sphere.

By comparing the panels at $t=t_{0}$ and at $t=4t_{0}$ in Figure \ref{fig:map_sT}, we see that,
as time goes by, the volume around the body
having low values of $|{\mathcal{T}}|$ becomes increasingly larger. 
At larger times and for the subsonic case, ${\mathcal{T}}$ asymptotically approaches to zero, 
i.e. the wake loses memory of the mass history,
except in a very narrow region in the very outer parts of the sonic sphere.
For supersonic perturbers, $|{\mathcal{T}}|$ decreases with time 
in most parts of the wake, both in the sonic sphere and within the Mach cone. 
In the limit $t\rightarrow \infty$,
$|{\mathcal{T}}|$ decreases at any point because of the geometrical dilution of the sound
waves launched by the perturber when it had a different mass. However, it
is interesting to note that even at $t=4t_{0}=8\tau$, 
when the perturber has almost reached its final mass,
${\mathcal{T}} >0.8$ in a significant portion of the wake for the case $\lambda=-0.75$, either 
if the body moves subsonically or supersonically.

Given $t$, $t_{0}$, $\tau$ and $\mu$, the shape of the contours of ${\mathcal{T}}$ does not 
depend on the value of $\lambda$ (see Figure \ref{fig:map_sT}). 
The reason is that the functions $\xi_{1}$, $\xi_{2}$
and $\xi_{3}$, which determine the shape of the contours, do not depend
on $\lambda$\footnote{Consider two models A and B with the same parameters $t_{0}$, $\tau$ and $\mu$
but different $\lambda$. For illustration, suppose that model A has $\lambda=-0.75$ and
model B has $\lambda=3$. We obtain  
${\mathcal{T}}_{B}(\vecx)=-0.39 {\mathcal{T}}_{A}(\vecx)$ at $t=2\tau$, and
 ${\mathcal{T}}_{B}(\vecx)=-0.25 {\mathcal{T}}_{A}(\vecx)$ at $t=8\tau$.}.

\section{Dynamical friction force }
\label{sec:dynamical_friction}
In this Section, the gravitational tug on the perturber by the overdense
wake is calculated using Equation (\ref{eq:DF_original}).
For a perturber in rectilinear orbit with velocity $V\hat{\vecz}$ and $V>0$, the gravitational drag 
can be recast in terms of $s=z-Vt$ and $R$ as
\begin{equation}
\vecF_{\rm DF}=2\pi GM_{p}\rho _{\infty}\hat{\vecz}\iint ds\ dR\frac{\alpha Rs}{\left(
s^{2}+R^{2}\right) ^{3/2}}.
\label{eq:force_cyl}
\end{equation}
We expect that $\vecF_{\rm DF}$ slows down the perturber even if the mass of the perturber varies over time.
Consequently, if we write $\vecF_{\rm DF}=-F_{\rm DF}\hat{\vecz}$, we expect that $F_{\rm DF}>0$.

At $-t_{0}<t<0$, the mass of the perturber
is constant and, therefore, the drag force is given in \citet{ost99}:
\begin{equation}
F_{\rm DF}=\frac{4\pi\rho_{\infty}(GM_{0})^{2}}{V^{2}} \ln \Lambda,
\label{eq:ostriker0}
\end{equation}
where $\ln\Lambda$ is given by Equation (\ref{eq:ostriker1}) for subsonic perturbers, and 
 \begin{equation}
\ln\Lambda=
\ln\left[\frac{V(t+t_{0})}{r_{\rm min}}\right]+\frac{1}{2}\ln \left(1-\mu^{-2}\right),
\label{eq:ostriker2}
\end{equation}
for supersonic perturbers ($\mu>1$).
These expressions for $\ln\Lambda$ are valid for $-t_{0}+r_{\rm min}/|V-c_{s}|<t<0$.

In the following, we wish to quantify $F_{\rm DF}$ at $t>0$, that is, when the episode of mass 
loss or mass gain has started. Since the linear analysis for a point-mass wake 
is only valid at $t> t_{\rm min}\equiv r_{\rm min}/|V-c_{s}|$ (see Appendix \ref{sec:appA} and 
Ostriker 1999), we will focus on estimating $F_{\rm DF}$ at $t>t_{\rm min}$.

To make comparisons, it is convenient to define $F_{\rm cst}$ as the strength of  the DF force 
using the instantaneous mass approximation, i.e. by assuming that the wake behind the perturber 
has density $\alpha_{\rm cst}$. More specifically,
$F_{\rm cst}$ is given by Equation (\ref{eq:ostriker0}), 
but replacing $M_{0}$ for $M_{p}(t)$. So,
\begin{equation}
F_{\rm cst}= {\mathcal{F}} \ln\Lambda,
\end{equation}
with
\begin{equation}
{\mathcal{F}}(t)= \frac{4\pi \rho_{\infty}(GM_{p}(t))^{2}}{V^{2}}.
\end{equation}
$\ln \Lambda$ is given by Equation (\ref{eq:ostriker1}) for subsonic perturbers, 
and by Equation (\ref{eq:ostriker2}) for supersonic perturbers.

We wish to determine how much the DF force $F_{\rm DF}$ deviates from $F_{\rm cst}$. 
To do so, we combine
Equations (\ref{eq:MpM0}), (\ref{eq:diff_alpha1}) and (\ref{eq:force_cyl}) to find that 
\begin{equation}
F_{\rm DF}(t)=F_{\rm cst}(t)+F_{\rm mem}(t),
\end{equation}
where the second term $F_{\rm mem}(t)$, which we will refer to it as the memory term, is given by
\begin{equation}
F_{\rm mem}= -\frac{\lambda }{\eta}  {\mathcal{F}} f_{\bullet},
\label{eq:Fmem1}
\end{equation}
where
\begin{equation} 
f_{\bullet}=f_{\bullet,1}+f_{\bullet,2}\exp(-t/\tau) 
\label{eq:bb}
\end{equation}
and
\begin{equation}
f_{\bullet,1}=\frac{\mu^{2}}{2}\iint \frac{(\xi_{2}-\xi_{1})Rs}{D\left(s^{2}+R^{2}\right)^{3/2}}\,ds\, dR,
\label{eq:fbullet_orig}
\end{equation}
\begin{equation}
f_{\bullet,2}=\frac{\mu^{2}}{2}\iint \frac{(\xi_{1}-\xi_{3})Rs}{D\left(s^{2}+R^{2}\right)^{3/2}}\,ds\, dR.
\label{eq:fbullet2}
\end{equation}
The integral in $f_{\bullet,1}$ can be performed analytically to obtain
\begin{equation}
f_{\bullet,1}(t)=\left\{ 
\begin{tabular}{l}
$0$\ \hskip 1.85cm if $\mu<1$ \\ 
$\ln \left(1+t_{0}/t\right)$\ \ \ \ if $\mu >1$.
\end{tabular}
\right.  
\label{eq:fbullet1}
\end{equation}
The two-dimensional integral $f_{\bullet,2}$ is much more complicated and will be computed 
numerically.

From Equation (\ref{eq:fbullet1}), it is obvious that $f_{\bullet,1}=0$ if $t_{0}=0$
(indeed $\xi_{1}=\xi_{2}$ in this case). In the general case, we have $f_{\bullet,1}\geq 0$.
On the other hand, even though $f_{\bullet,2}$ may be positive or negative,
we anticipate that $f_{\bullet}$ is always positive or zero, and therefore
$F_{\rm mem}$ is positive or negative depending on the sign of $\lambda$.
The memory term is positive when the perturber
loses mass ($\lambda<0$) because the wake is more dense than $\alpha_{\rm cst}$ (see \S \ref{sec:wake}).
Therefore, $F_{\rm cst}$ underestimates the drag force in this case.
If the perturber gains mass, the memory term is negative, and $F_{\rm cst}$
overestimates the drag force.

For constant-mass perturbers (either for $\lambda=0$ or for $\tau\rightarrow \infty$),
the memory term must be zero. In fact, if
$\lambda=0$, it holds from Equation (\ref{eq:Fmem1}) that
$F_{\rm mem}=0$ and hence $F_{\rm DF}=F_{\rm cst}$.
On the other hand, if $\tau\rightarrow \infty$, Eq. (\ref{eq:bb}) implies
$f_{\bullet}=f_{\bullet,1}+f_{\bullet,2}$. Moreover, $\xi_{3}=\xi_{2}$
(see Appendix \ref{sec:appB}) and according to Eqs. (\ref{eq:Fmem1}), (\ref{eq:fbullet_orig}) and (\ref{eq:fbullet2}),
$f_{\bullet}=F_{\rm mem}=0$.

$F_{\rm DF}$ is expected to be different from $F_{\rm cst}$ if the perturber had a different mass 
in the past. In the following, we study how $F_{\rm mem}$ depends on 
the mass history of the perturber and on time. 
Before dealing with the general case, we examine a scenario where
$\tau\rightarrow 0$, which corresponds to a very rapid change of mass,
in the next subsection.

\subsection{Dynamical friction for the case $\tau=0$}
\label{sec:df_tau_zero}
The case $\tau=0$ describes a situation where the perturber undergoes an instantaneous boost
in mass at $t=0$. Assuming that $t_{0}\neq 0$, the perturber's mass changes from  
$M_{0}$ to  $M_{f}\equiv (1+\lambda)M_{0}$. 
From Equations (\ref{eq:piece_wise}) and (\ref{eq:bb}) 
we have that $\eta=1+\lambda$, and $f_{\bullet}=f_{\bullet,1}$ at $t>t_{\rm min}$. 
Therefore, from Eqs. (\ref{eq:Fmem1}) and (\ref{eq:fbullet1}), the memory term is given by
\begin{equation}
F_{\rm mem}(t)=\left\{ 
\begin{tabular}{l}
$0$\ \ \ \ \ \ \ \ \ \ \ \hskip 2.0cm if $\mu<1$ \\ 
$-\frac{\lambda}{1+\lambda}{\mathcal{F}}_{0}\ln \left(1+t_{0}/t\right)$ \hskip 0.31cm if $\mu >1$,
\end{tabular}
\right.  
\label{eq:Fmem_tau0}
\end{equation}
with
\begin{equation}
{\mathcal{F}}_{0}= \frac{4\pi \rho_{\infty}(GM_{f})^{2}}{V^{2}}.
\end{equation}
Hence $F_{\rm mem}\rightarrow 0$ as $t\rightarrow \infty$.
One can show that $F_{\rm cst}+F_{\rm mem}>0$, implying that the net DF force is always opposed to 
the motion of the perturber.

\begin{figure}
\centering
 \includegraphics[width=0.475\textwidth]{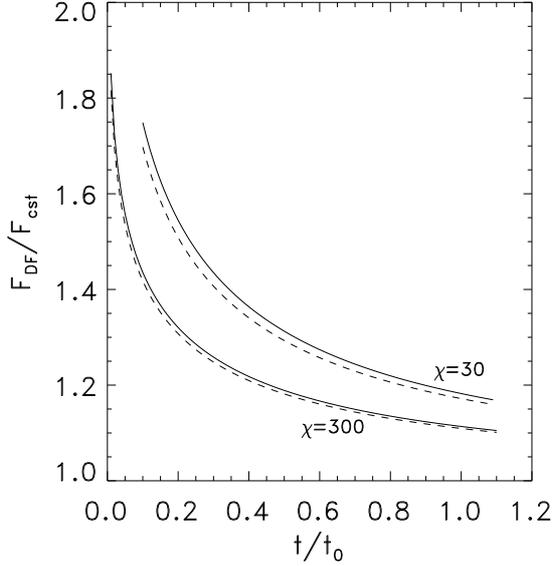}
 \caption{Ratio of $F_{\rm DF}$ to $F_{\rm cst}$ as a 
function of time, for $\tau=0$ and $\lambda=-0.5$, using Eq. (\ref{eq:Fmem_over_Fcst_tau0}). 
The solid lines correspond
to $\mu=1.5$ and the dashed lines for $\mu=3$.}     
 \label{fig:Fmem_tau0}
\end{figure}

According to Eq. (\ref{eq:Fmem_tau0}), $F_{\rm mem}=0$ if $\mu<1$.
This implies that the DF force
on a subsonic perturber with $\tau=0$ is identical to the DF force felt
by a perturber with constant mass $M_{f}$ since its birth.

For supersonic perturbers, at $t>t_{\rm min}$, the memory term in nonzero, except for $t_{0}=0$.
Interestingly, $F_{\rm mem}$ does not depend on $r_{\rm min}$. It does depend on
$\mu$ but just through the factor ${\mathcal{F}}_{0}$.
If $t_{0}=0$ (and $\tau=0$), $F_{\rm cst}$ gives the correct value of the force because the perturber 
has constant mass.

It is worthwhile to compare $F_{\rm mem}$ with $F_{\rm cst}$. 
For supersonic bodies, we find that
\begin{equation}
\frac{F_{\rm mem}}{F_{\rm cst}}
=-\frac{\lambda}{1+\lambda}\frac{\ln\left(1+t_{0}/t\right)}{\ln \Lambda}
\label{eq:Fmem_over_Fcst_tau0}
\end{equation}
where $\ln \Lambda$ is given in Equation (\ref{eq:ostriker2}). The instantaneous mass
approximation is satisfactory if $F_{\rm mem}/F_{\rm cst}$ is small.
In order to show the dependence of $F_{\rm mem}/F_{\rm cst}$ on time, we consider
cases where the length of the wake at $t=0$ is much larger than $r_{\rm min}$.
Since the length of the wake at $t=0$ is $\simeq Vt_{0}$,  we will focus on cases
where $\chi\equiv Vt_{0}/r_{\rm min}\gg 1$.
Figure \ref{fig:Fmem_tau0} shows $F_{\rm DF}/F_{\rm cst}$ (which it is just
$1+[F_{\rm mem}/F_{\rm cst}]$) as a function of time for 
$\lambda=-0.5$ and two different values of $\chi$. 
We see that $F_{\rm DF}/F_{\rm cst}$ decreases with time, and their values are
rather insensitive to the Mach number of the perturber.
At a given $t/t_{0}$, $F_{\rm DF}/F_{\rm cst}$ is larger for low values of $\chi$. 
At $t=t_{\rm min}$, $F_{\rm cst}$
underestimates the drag force by a factor of $\sim 1.8$. 

For any arbitrary $\lambda$,
the average value of $F_{\rm DF}/F_{\rm cst}$ between $t_{\rm min}$
and $t_{0}$, is $\simeq -1.32\lambda/(1+\lambda)$ for $\chi=30$, whereas it is 
$\simeq -1.23\lambda/(1+\lambda)$ for $\chi=300$.
Note that $t_{\rm min}$ depends on the value of $\chi$
as follows $t_{\rm min}=r_{\rm min}/|V-c_{s}|= [\mu/(\mu-1)]t_{0}/\chi$.

Under our assumption here that $\tau=0$, values for $\lambda$ more negative than $-0.5$ are 
not very realistic because the remnant would become gravitationally unbound and 
completely disrupted in a short timescale. 
For $\lambda\simeq -0.5$, $F_{\rm mem}$ is a sizeable fraction of $F_{\rm cst}$ on
a characteristic timescale $\sim t_{0}$.

\subsection{Dynamical friction for the case $\tau>0$}

In this subsection, we obtain $F_{\rm mem}$ for models with $\tau >0$.
In this general case, we need to compute  $f_{\bullet,2}$. This term takes into account
the contribution of the wake $\alpha_{3}$ (see Equation \ref{eq:fbullet2}).

\begin{figure}
\centering
 \includegraphics[width=0.475\textwidth,height=6.5cm]{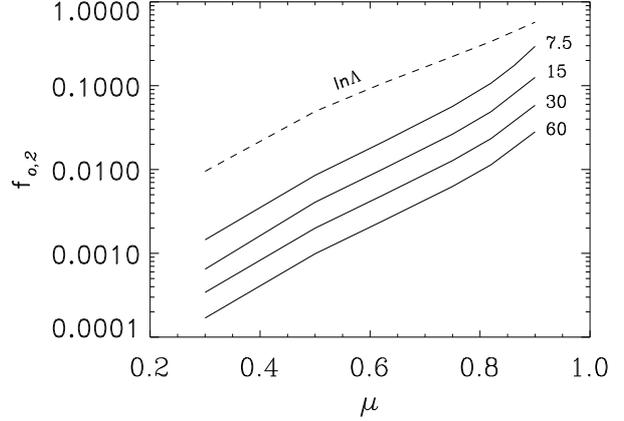}
 \caption{$f_{\bullet,2}$ versus Mach number for $\hat{\gamma}=7.5, 15, 30$
and $60$ (solid lines). For comparison, $\ln\Lambda$ is also plotted (dashed line).}     
 \label{fig:fstar2_subsonic}
\end{figure}

\subsubsection{Subsonic perturbers}
According to Eq. (\ref{eq:fbullet1}), $f_{\bullet,1}=0$ for subsonic perturbers.
Thus, 
\begin{equation}
F_{\rm mem}=-\left(\frac{\lambda}{\eta}\right) {\mathcal{F}}f_{\bullet,2}\exp(-t/\tau).
\end{equation}
$f_{\bullet,2}$ is obtained by integrating numerically Equation (\ref{eq:fbullet2}).
We find that $f_{\bullet,2}$ remains constant over time for subsonic perturbers.
More specifically, in the subsonic case, $f_{\bullet,2}$ only depends on two dimensionless parameters 
$\mu$  and $\hat{\gamma}$, which is defined as $\hat{\gamma}\equiv c_{s}\tau/r_{\rm min}$, i.e. 
\begin{equation} 
f_{\bullet,2}=f_{\bullet,2}\left(\mu,\hat{\gamma}\right).
\end{equation}
Note that $f_{\bullet,2}$ depends neither on $t$ nor on $t_{0}$ in the subsonic case.

Figure \ref{fig:fstar2_subsonic} shows $f_{\bullet,2}$ as a function of $\mu$,
for different values of $\hat{\gamma}$. The values of $f_{\bullet,2}$ are positive.
For a fixed value of $\hat{\gamma}$, $f_{\bullet,2}$ increases steeply with $\mu$.
On the other hand, for a given $\mu$, the function $f_{\bullet,2}$
and thereby $|F_{\rm mem}|$ decrease with $\hat{\gamma}$.
In the following, we estimate the contribution of $F_{\rm mem}$ to the 
drag force in order to find the values of $\hat{\gamma}$ for which the
instantaneous approximation could be adequate.

At $t>0$, the following inequality holds for any value of $\lambda$
\begin{equation}
|F_{\rm mem}|=\frac{|\lambda|}{\eta} {\mathcal{F}} f_{\bullet,2} \exp(-t/\tau)\leq
|\lambda| f_{\bullet,2}{\mathcal{F}}.
\end{equation}
On the other hand, we have that $F_{\rm cst}={\mathcal{F}}\ln\Lambda$.
Therefore, the ratio of $|F_{\rm mem}|$ to $F_{\rm cst}$
is $|\lambda| f_{\bullet,2}/\ln\Lambda$ at most.
In Figure \ref{fig:fstar2_subsonic}, we can 
compare $f_{\bullet,2}$ with $\ln\Lambda$. We
see that for $\mu<0.9$ and $\hat{\gamma}\gtrsim 15$,
$\ln\Lambda$ is a factor $\sim 5$ larger than $f_{\bullet,2}$. Therefore,
for those values of $\mu$ and $\hat{\gamma}$, 
the instantaneous approximation has a percentage error of at most $20\%$ for perturbers
with $|\lambda|\lesssim 1$. 

\begin{figure}
\centering
 \includegraphics[width=0.485\textwidth,height=12.2cm]{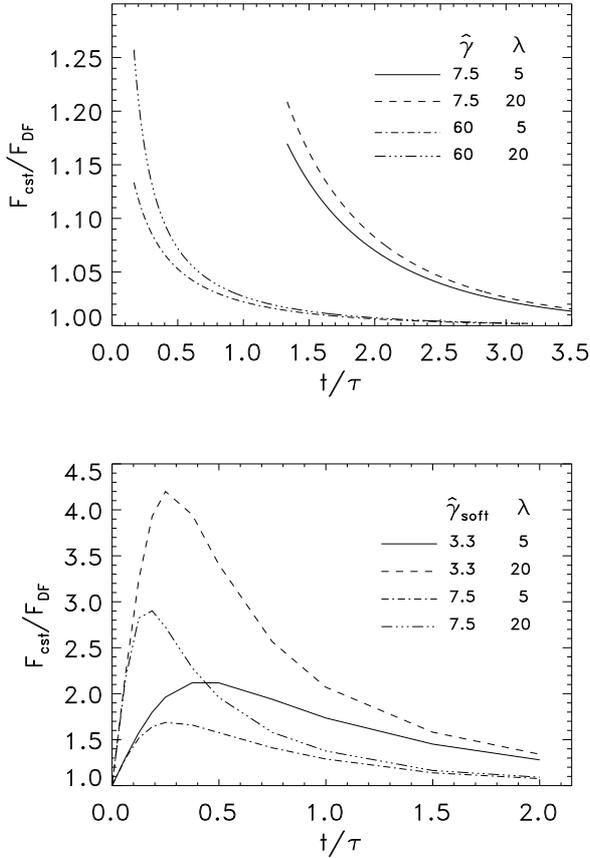}
 \caption{Evolution of the ratio $F_{\rm cst}/F_{\rm DF}$ 
for a point-mass perturber (top panel) and for a Plummer perturber (bottom panel),
for various values of $\hat{\gamma}$ and $\lambda$. In all cases $\mu=0.9$.
}     
 \label{fig:Fmem_subsonic}
\end{figure}

Consider now the case $\lambda>1$. The relative 
contribution of $F_{\rm mem}$ to the drag force increases with $|\lambda|$.
This is exemplified in Figure \ref{fig:Fmem_subsonic}, which
shows $F_{\rm cst}/F_{\rm DF}$ for $\mu=0.9$ and two different values of $\lambda$. 
For point-mass perturbers and for the values of $\hat{\gamma}$ explored in Figure \ref{fig:Fmem_subsonic}
(namely, $7.5\leq \hat{\gamma} \leq 60$),
$F_{\rm cst}$ overestimates the drag force by less than $25\%$, even for $\lambda=20$. 
We warn that this claim is valid at $t>t_{\rm min}$,
because linear theory cannot predict the drag force in the interval $0<t<t_{\rm min}$,
for point-mass perturbers.
It is worth noting that $t_{\rm min}$ in terms of $\mu$ and $\hat{\gamma}$ is
\begin{equation}
\frac{t_{\rm min}}{\tau}=\frac{1}{|1-\mu|\hat{\gamma}}.
\end{equation}
Therefore, for $\hat{\gamma}\leq 1/(1-\mu)$, linear theory
cannot predict the drag force in a time interval $\geq \tau$.

For softened perturbers, however, the response is linear at any location if the softening
radius $R_{\rm soft}$ is large compared to the accretion radius. In this
case, $F_{\rm DF}$ can be calculated at any time using $\alpha_{\rm soft}$ 
given in Equation (\ref{eq:softened_alpha}). In the bottom panel of Figure \ref{fig:Fmem_subsonic},
we plot $F_{\rm cst}/F_{\rm DF}$ for a Plummer perturber with radius $R_{\rm soft}$ and
moving at Mach number $\mu=0.9$.
Here $F_{\rm cst}$ denotes the drag force using $\alpha_{\rm cst}$ for 
a softened Plummer perturber. 
For a fair comparison between point-mass and extended perturbers, we have used 
$\hat{\gamma}_{\rm soft}=c_{s}\tau/r_{\rm min}=c_{s}\tau/(2.25R_{\rm soft})$
for Plummer perturbers instead of $\hat{\gamma}$
\citep[see][]{san99,ber13}.

In Figure \ref{fig:Fmem_subsonic} we see that $F_{\rm cst}/F_{\rm DF}$ reaches a
maximum and then declines with time, approaching $1$ asymptotically.
For $\mu=0.9$, $\lambda=5$ and $\hat{\gamma}_{\rm soft}=3.3$, the mean value of 
$F_{\rm cst}/F_{\rm DF}$ between
$t=0$ and $t=2\tau$ is $1.65$, whereas it is $2.4$ for $\lambda=20$ and
$\hat{\gamma}=3.3$. For $\mu<0.9$, $\lambda<5$ and $\hat{\gamma}>7.5$,
the above average value is less than $1.25$ (i.e., $F_{\rm cst}$ overestimates
the true drag force by $25\%$ or less).

\begin{figure}
\centering
 \includegraphics[width=0.485\textwidth,height=6.4cm]{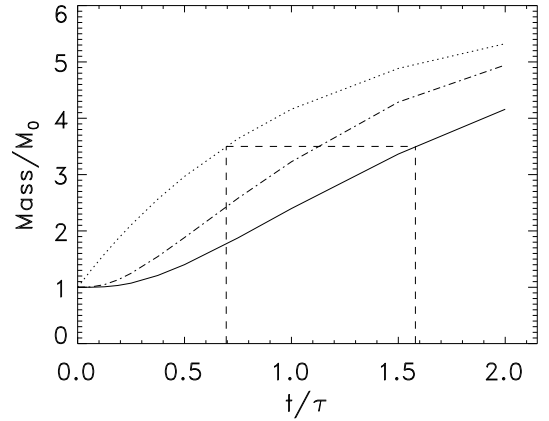}
 \caption{Effective mass $M_{\rm eff}/M_{0}$ for a model with $\mu=0.9$ and $\lambda=5$,
for $\hat{\gamma}_{\rm soft}=3.3$ (solid line) and for $\hat{\gamma}_{\rm soft}=7.5$ (dot-dash
line). For comparison we also plot the instantaneous mass $M_{p}/M_{0}$ (dotted line).
}     
 \label{fig:effective_mass_sub}
\end{figure}

The result that $F_{\rm DF}$ depends on $\hat{\gamma}_{\rm soft}$ implies that
if we wish to know the temporal evolution of $F_{\rm DF}$, we need to know
how the mass is added to the perturber. For instance, if a Plummer body
grows in mass keeping constant its central density, its radius should
increase over time according to the law $R_{\rm soft}(t)=R_{\rm soft}(0)\eta^{1/3}$.
As a consequence, $\hat{\gamma}_{\rm soft}$ declines with time. This will result in a more
rapid drop of $F_{\rm cst}/F_{\rm DF}$ over time. This effect will be more notorious
for large values of $\lambda$.

The rate of change of the velocity of the perturber is $F_{\rm DF}/M_{p}$.
The classical Ostriker formula can be used to estimate this frictional
deceleration if the mass of the perturber is formally replaced by an
effective mass $M_{\rm eff}(t)$. By definition, $M_{\rm eff}$ satisfies the relationship
\begin{equation}
\frac{F_{\rm DF}}{M_{p}}= \frac{4\pi \rho_{\infty} G^{2}M_{\rm eff}}{V^{2}} \ln \Lambda.
\end{equation}
Figure \ref{fig:effective_mass_sub} plots the effective mass for models with
$\mu=0.9$ and $\lambda=5$. For positive $\lambda$, the effective mass is always
smaller than the instantaneous mass. We see that for the model with 
$\hat{\gamma}_{\rm soft}=3.3$, the effective mass at $t=1.6\tau$ equals the 
perturber's mass at $t=0.7\tau$ (dashed line). In fact, the curves of the effective mass with time
are less steep than the curve for the instantaneous mass.

\begin{figure}
\centering
 \hbox{\hspace{-1.7em}\includegraphics[width=0.5\textwidth,height=17.0cm]{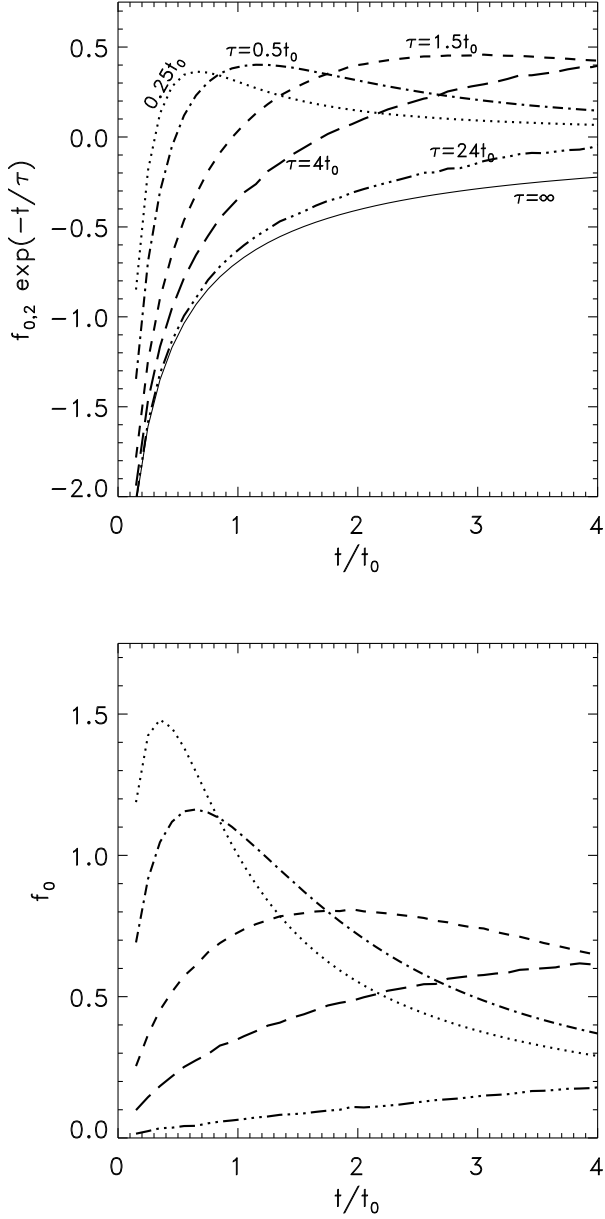}}
 \caption{The evolution of the terms $f_{\bullet,2}\exp(-t/\tau)$ (top panel)
and $f_{\bullet}$ (bottom panel) for a supersonic perturber with $\mu=1.5$ 
and $\chi=30$. Curves for different values of $\tau$, as labelled on each
curve, are shown.
}     
 \label{fig:fstar2_supersonic30}
\end{figure}

\subsubsection{Supersonic perturbers}
In this subsection we consider bodies in supersonic motion $\mu>1$. 
Combining Equations (\ref{eq:bb}) and (\ref{eq:fbullet1}), we find
\begin{equation}
f_{\bullet}=\ln\left[\frac{t_{0}+t}{t}\right]+f_{\bullet,2}\exp(-t/\tau).
\label{eq:Fmem_supersonic}
\end{equation}
In the limit $\tau\rightarrow 0$,
the second term in the right-hand-side of Equation (\ref{eq:Fmem_supersonic}) can be neglected,
and we recover the case discussed in Section \ref{sec:df_tau_zero}.

In the following, we take $t_{0}\neq 0$, unless metioned otherwise.
The function $f_{\bullet,2}$ depends on four dimensionless parameters:
$t/t_{0}, \mu,\tau/t_{0}$ and $\chi$.
Recall that $\chi$ was defined in Section \ref{sec:df_tau_zero} as $\chi=Vt_{0}/r_{\rm min}$.
The value of $f_{\bullet,2}$ can be found analytically in the limit $\tau\rightarrow \infty$ and it is
given by 
\begin{equation}
f_{\bullet,2}=\ln \left(\frac{t}{t_{0}+t}\right)=\ln\left(\frac{t/t_{0}}{1+t/t_{0}}\right).
\end{equation}
Note that $f_{\bullet,2}<0$ in this limit.

Figure \ref{fig:fstar2_supersonic30} shows $f_{\bullet,2}\exp(-t/\tau)$ and $f_{\bullet}$ versus time 
for $\mu=1.5$ and $\chi=30$. 
For these values of $\mu$ and $\chi$,
$t_{\rm min}/t_{0}=0.1$. We have used that $t_{\rm min}/t_{0}$ can be
recast in terms of $\mu$ and $\chi$ as
\begin{equation}
\frac{t_{\rm min}}{t_{0}} =\frac{\mu}{(\mu-1)\chi}.
\end{equation}
Both $f_{\bullet,2}\exp(-t/\tau)$ and $f_{\bullet}$ achieve a maximum
and then decrease with time, approaching to zero at large $t$.
The curves with $\tau=4t_{0}$ and $\tau=24t_{0}$ reach the maximum
at $t>4t_{0}$, outside the range displayed in Figure \ref{fig:fstar2_supersonic30}. 
The peak value of $f_{\bullet}(t)$ decreases as $\tau/t_{0}$ increases.
For instance, the maximum value of $f_{\bullet}$ is $1.5$ for $\tau=0.5t_{0}$,
whereas it is $0.65$ for $\tau=4t_{0}$.

\begin{figure}
\centering
 \includegraphics[width=0.5\textwidth,height=7.4cm]{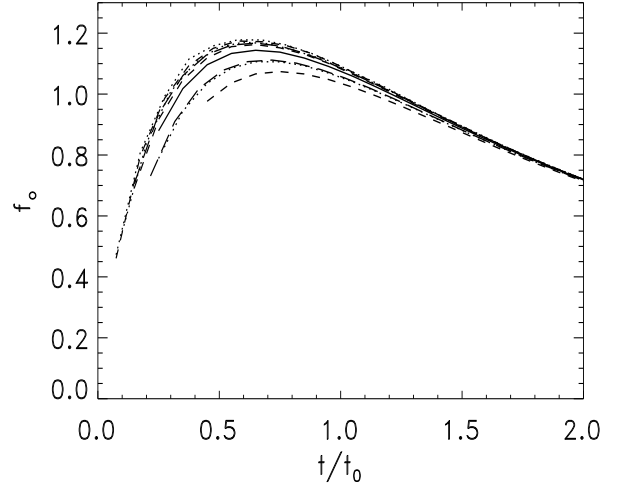}
 \caption{$f_{\bullet}$ as a function of time for various combinations of $\mu$ and $\chi$. 
In all cases $\tau=0.5t_{0}$. From top to bottom: $(\mu,\chi)= (3, 60), (1.5, 60), (1.2, 60),
(3, 30), (1.5,30), (1.2, 30), (5, 7.5), (3, 7.5)$ and $(1.5, 7.5)$.
}     
 \label{fig:fstar2_diff_chi}
\end{figure}

From the top panel of Figure \ref{fig:fstar2_supersonic30}, we see that 
$f_{\bullet,2}\exp(-t/\tau)\geq \ln(t/[t_{0}+t])$, which implies that $f_{\bullet}>0$
as the bottom panel shows.
Thereby $F_{\rm mem}$ is positive if $\lambda<0$, and it is negative if $\lambda>0$
(see Equation \ref{eq:Fmem1}). 

Figure \ref{fig:fstar2_diff_chi} shows $f_{\bullet}$ for $\tau=0.5t_{0}$, 
varying  $\mu$ and $\chi$. It is seen that $f_{\bullet}$ depends
weakly on $\mu$ and $\chi$; changing $\mu$ by a factor of $3$
and $\chi$ by a factor of $8$, $f_{\bullet}$ varies less than $10\%$.
For $\tau=4t_{0}$, this variation is less than $16\%$ (not shown).
This implies that the precise values of $\mu$ and $r_{\rm min}$ are not crucial to estimating
$f_{\bullet}$. 
Indeed, we find that when $\lambda$ is given, 
$F_{\rm mem}/F_{\rm cst}$ is rather insensitive to $\mu$ and $\chi$, provided they lie
in the range $\mu>1.2$ and $\chi>7.5$.
For this reason, we will focus our discussion on the case $\mu=1.5$ and $\chi=30$, but all
the results are also valid in the abovementioned range of $\mu$ and $\chi$.

\begin{figure}
\centering
\hbox{\hspace{-1.7em} \includegraphics[width=0.5\textwidth,height=13.4cm]{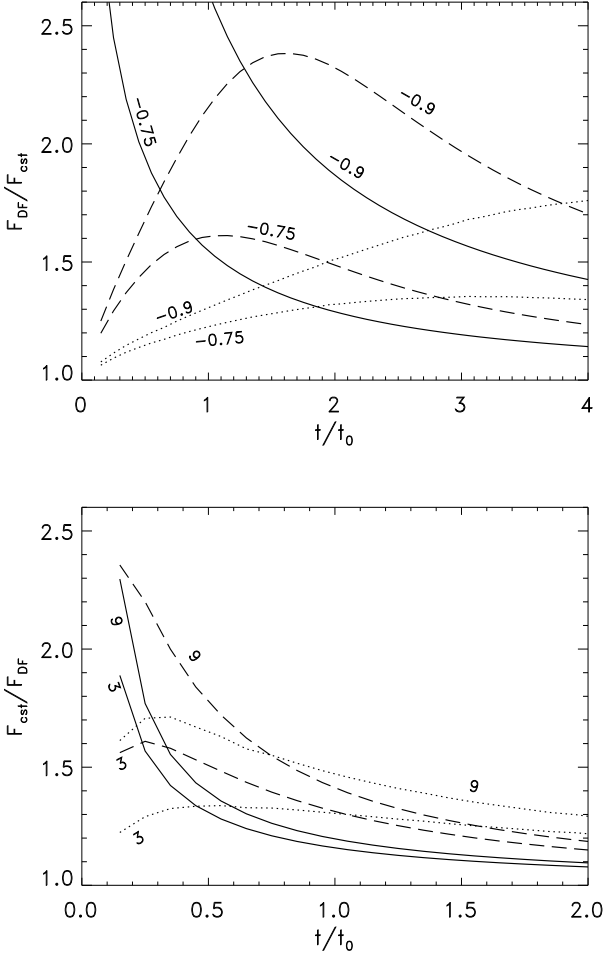}}
 \caption{The evolution of $F_{\rm DF}/F_{\rm cst}$ (top panel) and $F_{\rm cst}/F_{\rm DF}$
(bottom panel) for a supersonic perturber
with $\mu=1.5$ and $\chi=30$, and different combinations of $\lambda$ and
$\tau$. 
Solid lines correspond to $\tau=0$, dashed lines for $\tau=0.5t_{0}$ 
and dotted lines for $\tau=1.5t_{0}$. The values of $\lambda$ are labelled
on each curve.
}     
 \label{fig:Fmem_over_Fcst_supersonic}
\end{figure}

In order to quantify how much $F_{\rm cst}$ departs from the true drag force $F_{\rm DF}$,
Figure \ref{fig:Fmem_over_Fcst_supersonic} displays $F_{\rm DF}/F_{\rm cst}$  
for some representative cases, using $\mu=1.5$ and $\chi=30$. Let us focus first on 
the case $\lambda=-0.9$. For $\tau=0$, $F_{\rm DF}/F_{\rm cst}$ decreases
rapidly and continuously with time, approaching $1$. 
At $t<1.5t_{0}$, it is larger than $2$, implying that $F_{\rm cst}$ underestimates
the drag force by a factor larger than $2$. 
For $\tau=0.5 t_{0}$ (and $\lambda=-0.9$), $F_{\rm DF}/F_{\rm cst}$ rises
from $t_{\rm min}$ to $1.6t_{0}$, where it reaches a maximum and then declines.
It is larger than $2$ between $0.8t_{0}$ and $3t_{0}$. Finally, for $\tau=1.5t_{0}$,
$F_{\rm DF}/F_{\rm cst}$ gradually increases along the time interval displayed in Figure \ref{fig:Fmem_over_Fcst_supersonic}, but it is always below $2$. 

\begin{figure}
\centering
\hbox{\hspace{-1.7em} \includegraphics[width=0.5\textwidth,height=13.4cm]{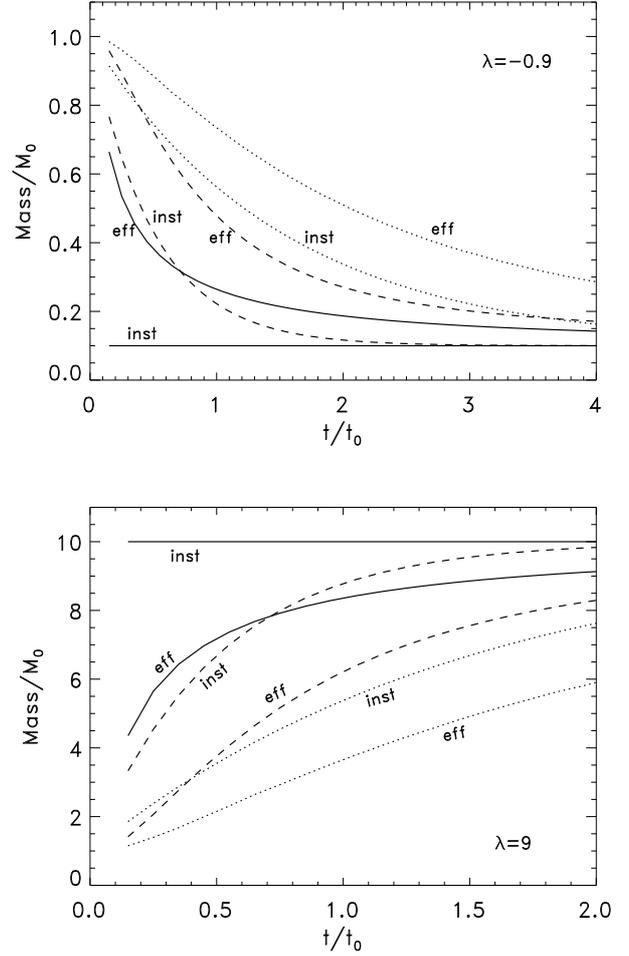}}
 \caption{Effective mass (curves marked with the label `eff') and instantaneous mass (curves labelled with
`inst') as a function of time, for a perturber with $\mu=1.5$ and $\chi=30$.
We show cases with $\lambda=-0.9$ (top panel) and $\lambda=9$ (bottom panel).
As in Figure \ref{fig:Fmem_over_Fcst_supersonic}, solid lines correspond to $\tau=0$, 
dashed lines for $\tau=0.5t_{0}$ and dotted lines for $\tau=1.5t_{0}$. 
}     
 \label{fig:effective_mass_sup}
\end{figure}

In the interval $t_{\rm min}<t<t_{0}$, the average value of $F_{\rm DF}/F_{\rm cst}$
is $4.0$, $1.7$ and $1.2$ for $\tau=0$, $0.5$ and $1.5$, respectively.
In the interval $t_{\rm min}<t<4t_{0}$, the average values are $2.3$, $2.0$ and
$1.5$, respectively.
It is remarkable that for a perturber with $\tau=0.5t_{0}$,
having almost reached its final mass at $t=2t_{0}$, the true drag force is still 
as large as $\sim 1.7 F_{\rm cst}$ at $t=4t_{0}$.

For $\lambda=-0.75$, the curves $F_{\rm DF}/F_{\rm cst}$ versus time behave
rather similar to $\lambda=-0.9$, but with less amplitude. For instance, 
for $\tau=0.5t_{0}$, $F_{\rm DF}$ is always less than $1.65 F_{\rm cst}$.
In other words, $F_{\rm mem}$ contributes to the drag force $65\%$ or less.
For $\tau=1.5 t_{0}$, $F_{\rm mem}$ contributes less than $40\%$.

For mass-gaining objects with $\lambda=9$, the average values of $F_{\rm cst}/F_{\rm DF}$ 
between $t_{\rm min}$ and $t_{0}$ are $1.45$, $1.75$ and $1.6$ for $\tau=0$, $0.5t_{0}$ and 
$1.5t_{0}$, respectively.
At $t>t_{0}$, $F_{\rm cst}$ overpredicts $F_{\rm DF}$ by $50\%$ or less,
even for $\lambda=9$. In particular, at $t=4t_{0}$, $F_{\rm cst}/F_{\rm DF}=1.1$.
If we compare the curves for $\lambda=9$
with those for $\lambda=-0.9$ (in both cases the mass of the perturber changes
by a factor of $10$), we find that the memory effect is longer-lasting for $\lambda=-0.9$.

The effective mass for some of these models is shown in Figure \ref{fig:effective_mass_sup}.
The offset between $M_{\rm eff}$ and $M_{p}$ is due to the history dependent nature
of the process. As a consequence, the effective mass always
takes a value between $M_{0}$ and $M_{p}$.

\begin{figure}
\centering
 \includegraphics[width=0.5\textwidth,height=11.8cm]{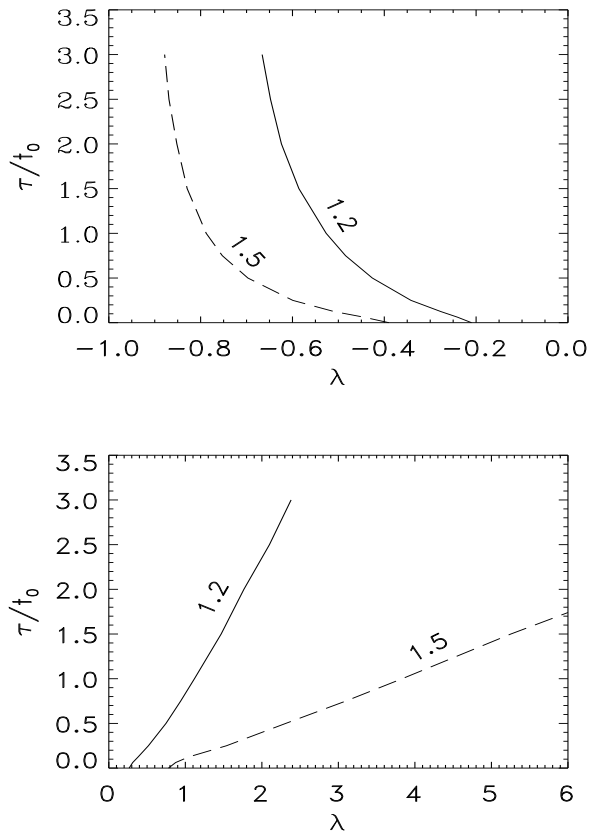}
 \caption{Diagram to illustrate the accuracy of $F_{\rm cst}$ for estimating the drag force.
Below the curves, $F_{\rm cst}$ underestimates (for $\lambda<0$) or overestimates ($\lambda>0$)
the drag force by a factor larger than $1.2$ (solid lines) or by a factor greater than $1.5$ (dashed lines).
We have used $\mu=1.5$ and $\chi=30$.
}     
 \label{fig:save_zone}
\end{figure}

Figure \ref{fig:save_zone} shows a diagram in the ($\lambda$, $\tau$) plane, which
outlines the parameter space where $F_{\rm cst}$ predicts correctly the value of the
drag force within a factor of $1.2$ at any time. To build the plot, we use $\mu=1.5$
and $\chi=30$.
For instance, for $\lambda=-0.6$, the error made by using $F_{\rm cst}$ is less 
than $20\%$ if $\tau>1.5t_{0}$.
Since the curve is very stiff at $\lambda=-0.6$, we require $\tau>3.5 t_{0}$ to have the
same error for $\lambda=-0.7$. Still, for $\lambda=-0.7$ and $\tau>0.5t_{0}$, the error
is less than $50\%$. On the other hand, for positive values of $\lambda$, the condition
$F_{\rm cst}/F_{\rm DF}<1.2$ is much more stringent than the condition 
 $F_{\rm cst}/F_{\rm DF}<1.5$. For $\lambda=6$ and $\tau>1.7t_{0}$, $F_{\rm cst}$
overestimates $F_{\rm DF}$ by a factor less than $1.5$ at any time.

\begin{figure}
\centering
 \includegraphics[width=0.5\textwidth,height=6.5cm]{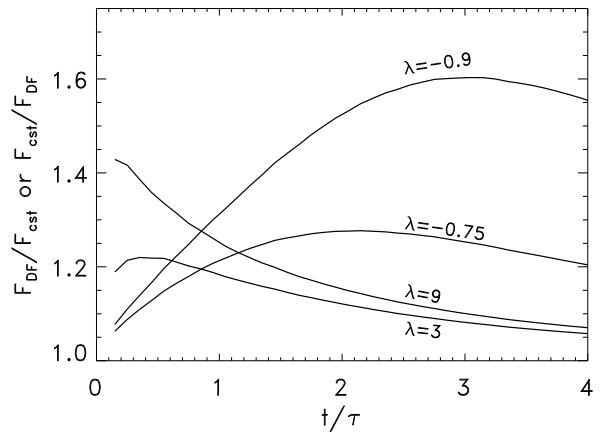}
 \caption{Time dependence of $F_{\rm DF}/F_{\rm cst}$ (if $\lambda<0$) and
$F_{\rm cst}/F_{\rm DF}$ (if $\lambda>0$) for $t_{0}=0$, $\mu=1.5$ and $\chi=30$,
and various values of $\lambda$. 
}     
 \label{fig:t0_zero}
\end{figure}

For completeness, Figure \ref{fig:t0_zero} compares $F_{\rm cst}$ with $F_{\rm DF}$ for $t_{0}=0$
and $V\tau/r_{\rm min}=30$. For $\lambda$ in the range $-0.9<\lambda <9$, they never differ 
from one another by a factor greater than $1.6$.

\section{Summary and final remarks}
\label{sec:summary}
We have studied the gravitational response of a gaseous medium to the gravitational pull
exerted by a body of time-varying mass.
To model this, we examine a scenario where the perturber has constant mass
during a lapse of time $t_{0}$, and immediately thereafter it 
undergoes an episode of gain or loss of mass with a characteristic timescale $\tau$.
We have assumed that the body moves
in rectilinear orbit through an initially homogeneous medium
The main goals were to characterize the density perturbation in the gas and to 
calculate how the DF force depends on the mass history of the perturber. 
Our main results can be summarized as follows:

(i) The structure of the wake contains a record
of the history of mass of the perturber.
For perturbers that lose mass, the wake is more dense than predicted using
the instantaneous mass in the classical Ostriker's formula, especially in the outer parts of the wake,
because the perturber was more massive in the past and thereby induced a stronger 
perturbation in the gas. If the perturber gains mass, the wake is less dense than predicted
in the instantaneous approximation.

(ii)  We have presented cases where
large regions of the wake still retain memory of the past mass of the
perturber even after a time $\sim 8\tau$, i.e. well after the episode of mass change has been
completed.

(iii) No matter the perturber mass history, the net DF force is always opposed to the 
perturber's velocity.

(iv) The DF force can be split into two components: the force predicted in
the instantaneous approximation $F_{\rm cst}$, plus a memory term.
The memory term is positive, i.e. it contributes to the drag force, if the 
perturber loses mass. It is negative if the perturber increases its mass.

(v) The standard formulas for the rate of change of the perturber's velocity
can be applied once the mass of the perturber is replaced with an
effective mass. We have shown that the effective mass is larger (smaller) than the 
instantaneous mass if the body loses (gains) mass.


The response of the gas depends on whether the body moves subsonically
or supersonically.
For {\it subsonic perturbers} of varying mass, we have found that

(i) The isodensity contours are not longer ellipsoids, and
the well-known front-back symmetry of the density about constant-mass subsonic
perturbers is broken down when the perturber's mass is not constant.

(ii) The memory term vanishes if the change in mass occurs instantly ($\tau=0$),
implying that $F_{\rm cst}$ accurately gives the DF force. 
If the change of mass occurs during a non-vanishing time interval, the 
relative contribution of the memory term increases with increasing Mach number.

(iii) We have provided examples of mass-gaining perturbers in which $F_{\rm cst}$ 
overestimates the drag force by more than $50\%$ on the interval between $t=0$
and $t\simeq 2\tau$, i.e. throughout the episode of mass loss/gain.

For {\it perturbers moving supersonically}, we have found that 

(i) The two critical parameters that determine the importance of the
memory term are the fractional change of
mass and the value of $\tau$ relative to $t_{0}$. The relative difference between 
$F_{\rm cst}$ and $F_{\rm DF}$ reduces as the ratio $\tau/t_{0}$ increases.

(ii) For instant gain or loss of mass ($\tau=0$), the memory term is initially 
large and then decreases with time to zero. For $\tau\neq 0$, the memory term
acquires its maximum value at $(0.5-2)\tau$.

(iii) As a rule-of-thumb, $F_{\rm cst}$ gives the drag force with a percent
error less than $20\%$ as long as the perturber's mass changes by less than $50\%$
and $\tau>1.5t_{0}$.

Our results should find application in a number of astrophysical settings,
such as the motion of dissolving systems (stellar clusters, globular
clusters or dwarf satellite galaxies), through interstellar gas of the parent galaxy.
Episodes of mass loss on short timescales may occur at pericentre passages or at
late stages of tidal disruption.

It is also worth mentioning that the e-folding timescale
for mass growth of stars and black holes embedded in a typical interstellar
cloud or nuclear disks, through Bondi-Hoyle-Lyttleton accretion, is much longer than the 
orbital timescale. Under these circumstances and based on our results,
the instantaneous approximation should be satisfactory.
In star-forming molecular clouds, however, massive stars may form by competitive
accretion in a timescale of the order of the free-fall time \citep[e.g.,][]{bon06}.
Nevertheless, the assumption that the unperturbed medium is
homogeneous and at rest is not strictly applicable due to the chaotic and turbulent
nature of the environment.

In this work, we have considered only the gravitational drag exerted on the
perturber from the density enhancement in the wake. In the case of a body that
gains mass, it can also experience an aerodynamic drag due to the direct
transfer of momentum by accreting material. 
For supersonic perturbers, the aerodynamic drag could be even
more important than the dynamical drag.

We have adopted some idealized assumptions to explore the DF acting
on perturbers of non-constant mass. In particular, we have assumed that
the perturber moves in rectilinear orbit. In real systems, however, the
orbit has some curvature. For constant-mass supersonic perturbers on an orbit with a
typical size $R_{0}$, the curvature of the orbit 
causes that the maximum impact parameter will cease to increase linearly with time.
For perturbers with $R_{\rm soft}\ll R_{0}/\mu$, the bending of the wake along 
the orbit leads to a cut-off in the impact parameter above
$(0.5-1)R_{0}$ \citep{san01,kim07,jus11}.
In this context, the time $t_{0}$ can be related to the orbital parameters by
equating $Vt_{0}\simeq (0.5-1)R_{0}$ and, therefore, $t_{0}$ should be identified with the 
orbital timescale $\approx (0.5-1)R_{0}/V$. On the other hand,
because of the curvature of the orbit, the
wake ``restarts'' over each one-quarter orbit. This leads to
a reduction of the memory term at times $\gtrsim R_{0}/V$
relative to the rectilinear orbit. Therefore, the estimates of the
memory term derived in the straight-line orbit 
should be considered as upper limits at $t\gtrsim R_{0}/V$.
The above reasoning is only valid if $R_{\rm soft}\ll R_{0}/\mu$.
If this condition is not fulfilled, the gravitational pull of the wake ahead of the
perturber cannot be ignored \citep[e.g.,][]{san18}, and a more delicate analysis is required.

\section*{Acknowledgments}
We thank the referee for constructive criticisms and valuable comments.
We also thank Ana Hidalgo for useful input.
The authors acknowledge funding from PAPIIT project IN111118.

\appendix

\section{Perturbed density profiles for wakes $1$ and $2$}
\label{sec:appA}
In Section \ref{sec:formal_solution}, we define $\alpha_{1}$ as the wake produced by a perturber 
with a density profile
\begin{equation}
\rho_{p,1}(\vecx,t)=M_{0}\delta(x)\delta(y)\delta(z-Vt)\Theta(t+t_{0}).
\end{equation}
On the other hand, $\alpha_{2}$ is the density perturbation in the wake induced by a body with 
a density distribution
\begin{equation}
\rho_{p,2}(\vecx,t)=\lambda M_{0}\delta(x)\delta(y)\delta(z-Vt)\Theta(t).
\end{equation}
It is straightforward to obtain $\alpha_{1}(\vecx,t)$ and $\alpha_{2}(\vecx,t)$ from Ostriker (1999).
In the case of $\rho_{p,1}$, the perturber is turned on at the location $(0,0,-Vt_{0})$.
At time $t<-t_{0}$, it holds $\alpha_{1}=0$. Now consider $t>-t_{0}$.
We define Region 1 as the sonic sphere centred at $(0,0,-Vt_{0})$:
\begin{equation}
R^{2}+(z+Vt_{0})^{2}< c_{s}^{2}(t+t_{0})^{2}
\end{equation} 
For subsonic perturbers, the wake $1$ is confined to Region 1.
For supersonic perturbers, we define Region 1$^{\prime}$ as the rear Mach cone,
which is the volume that satisfies the following conditions:
\begin{equation}
\frac{s}{R}<-(\mu^{2}-1)^{1/2},
\end{equation}
\begin{equation}
R^{2}+(z+Vt_{0})^{2}> c_{s}^{2}(t+t_{0})^{2},
\end{equation} 
and
\begin{equation}
z> \frac{c_{s}}{\mu}(t+t_{0})-\mu c_{s}t_{0}.
\end{equation}
At $t>-t_{0}$, the associated density wake $\alpha_{1}(\vecx,t)$ takes the form
\begin{equation}
\alpha_{1}(\vecx,t)=\frac{\xi_{1}G M_{0}}{c_{s}^{2}D},
\label{eq:alpha1appA}
\end{equation}
where $\xi_{1}=1$ in Region 1, $\xi_{1}=2$ in Region 1$^\prime$ only if $\mu>1$, and 
$\xi_{1}=0$ otherwise.

\begin{figure}
\centering
\hbox{\hspace{-2.3em}
 \includegraphics[width=0.55\textwidth,height=9.05cm]{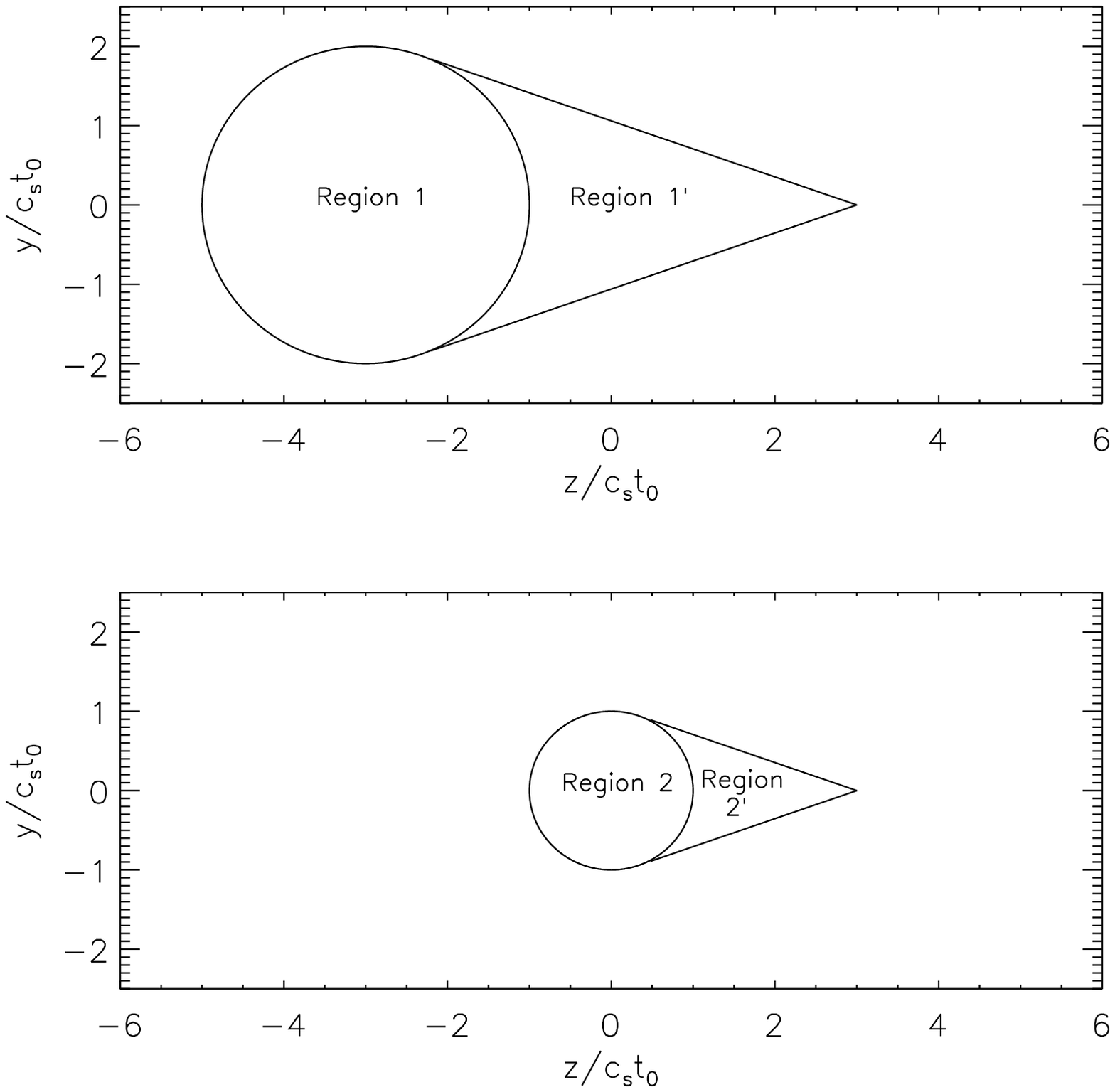}}
 \caption{The different regions defined in Appendix \ref{sec:appA}, at $t=t_{0}$,
for a supersonic perturber with $\mu=3$. For a supersonic perturber,
regions $1$ and $2$ make contact at $t=(\mu-1)t_{0}/2$.}     
 \label{fig:regions}
\end{figure}

The perturbed density $\alpha_{2}$ is zero at $t<0$. At $t>0$, it is 
\begin{equation}
\alpha_{2}(\vecx,t)=\frac{\xi_{2}G \lambda M_{0}}{c_{s}^{2}D},
\end{equation}
where $\xi_{2}=1$ in Region $2$, which is defined as the following sonic sphere
\begin{equation}
R^{2}+z^{2}< c_{s}^{2}t^{2}.
\end{equation} 
For supersonic perturbers ($\mu>1$), $\xi_{2}=2$ in Region $2^{\prime}$, which is specified by
these three inequalities
\begin{equation}
\frac{s}{R}<-(\mu^{2}-1)^{1/2},
\end{equation}
\begin{equation}
R^{2}+z^{2}> c_{s}^{2}t^{2},
\end{equation} 
and
\begin{equation}
z> \frac{c_{s}t}{\mu}.
\end{equation}
Outside regions $2$ and $2'$, the medium remains unperturbed and, consequently, $\xi_{2}=0$.
A representative diagram of the Regions $1$, $1'$, $2$ and $2'$ is 
shown in Figure \ref{fig:regions} for both subsonic and supersonic perturbers.

We recall that the above expressions for $\alpha_{1}$ and $\alpha_{2}$  are valid only for 
$D\gg GM_{0}/c_{s}^{2}$.

\section{Perturbed density for wake $3$}
\label{sec:appB}
As said in Section \ref{sec:formal_solution}, $\alpha_{3}$ denotes the density wake excited
by a fictitious perturber with the following mass density:
\begin{equation}
\rho_{p,3}(\vecx,t)=-\lambda M_{0} \exp (-t/\tau)\Theta(t)\delta(x)\delta(y)\delta(z-Vt).
\end{equation}
At $t>0$, $\alpha_{3}(\vecx,t)$ is obtained from 
Equation (\ref{eq:linearalpha}). After performing the integration 
over $t^{\prime}$, $x^{\prime}$ and $y^{\prime}$, which are trivial,
only the integral over $z^{\prime}$ is left. Using the variable
$\omega\equiv z-z^{\prime}$ instead of $z^{\prime}$ yields
\begin{eqnarray}
\alpha_{3}(R,z,t)=-\frac{\lambda G M_{0}}{c_{s}^{2}}
\int_{-\infty}^{\infty} d\omega \,\delta\left(\omega+s+\mu [R^{2}+\omega^{2}]^{1/2}\right)\nonumber\\ 
\times\exp\left(-\frac{\omega+z}{V\tau}\right)\frac{\Theta(\omega+z)}{(R^{2}+\omega^{2})^{1/2}}.
\end{eqnarray} 
To evaluate this integral, we find the two simple
zeros, $\omega_{+}$ and $\omega_{-}$, of the argument of the delta function:
\begin{equation}
\omega_{\pm}=\frac{s\pm \mu [s^{2}+R^{2}(1-\mu^{2})]^{1/2}}{\mu^{2}-1},
\end{equation}
and apply the identity that
\begin{equation}
\delta(f(\omega))=\sum_{j} \frac{\delta(\omega-\omega_{j})}{|f'(\omega_{j})|},
\end{equation}
where $\omega_{j}$ are the simple roots of $f(\omega)$.

As occurs in the derivation of $\alpha$ by a constant-mass perturber \citep{ost99},
only the root $\omega_{+}$ is valid if $\mu<1$.
In fact, $\omega_{-}$ is not a valid root if $\mu<1$ because it does not satisfy the
equality $\mu (R^{2}+\omega_{-}^{2})^{1/2}=-(\omega_{-}+s)$.
Therefore, if $\mu<1$, we have that
\begin{equation}
\alpha_{3}(\vecx,t)=-\frac{\lambda GM_{0}}{c_{s}^{2}D} \Theta(z+\omega_{+})
\exp\left(-\frac{t}{\tau}-\frac{\mu s + D}{(\mu^{2}-1) c_{s}\tau}\right).
\end{equation}
Given that $z+\omega_{+}$ is positive only if $R^{2}+z^{2}<c_{s}^{2}t^{2}$ (i.e. in Region
$2$), we find that 
\begin{equation}
\alpha_{3} \left( \vecx, t\right) =-\frac{\lambda GM_{0}}{c_{s}^{2} D}  
\exp \left[ -\frac{t}{\tau }+\frac{\mu s+D}{\left( 1-\mu
^{2}\right) c_{s}\tau }\right],
\end{equation}
in Region $2$, and it is zero outside region $2$, provided that $\mu<1$.

On the other hand, if $\mu>1$, both $\omega_{+}$ and $\omega_{-}$ are valid in the region
$s/R<-(\mu^{2}-1)^{1/2}$. Therefore, for $\mu>1$,
\begin{equation}
\alpha_{3}=-\frac{\lambda GM_{0}}{c_{s}^{2}D} \sum_{j=+,-}\Theta(z+\omega_{j})
\exp\left(-\frac{t}{\tau}-\frac{\mu s \pm D}{(\mu^{2}-1)c_{s}\tau}\right)
\label{eq:alpha3_supersonic}
\end{equation}
if $s/R<-(\mu^{2}-1)^{1/2}$, while $\alpha_{3}=0$ otherwise.

Now consider the argument of the Heaviside function in Equation (\ref{eq:alpha3_supersonic}). 
In Region $2'$ (and for $\mu>1$), it is easy to show that $z+\omega_{+}>0$ and $z+\omega_{-}>0$ and, 
therefore, $\Theta=1$ and both terms in the sum in the RHS of Equation (\ref{eq:alpha3_supersonic}) 
are nonzero. More specifically, the perturbed density of the wake $3$ in Region $2'$ is given by
\begin{eqnarray}
\nonumber\alpha_{3}  =-\frac{\lambda GM_{0}}{c_{s}^{2}D}  \exp\left(-\frac{t}{\tau}\right)\\ 
 \times\left\{ \exp \left[-\frac{\mu s+D}{\left(\mu^{2}-1\right)c_{s}\tau}\right] 
+\exp \left[-\frac{\mu s-D}{\left( \mu ^{2}-1\right)c_{s}\tau }\right] \right\}.
\end{eqnarray}

In Region $2$ (and for $\mu>1$), $z+\omega_{+}>0$ and $z+\omega_{-}<0$ and therefore only
the root $\omega_{+}$ contributes to the sum in Equation (\ref{eq:alpha3_supersonic}). In this
region,
\begin{equation}
\alpha_{3} \left( \vecx, t\right) =-\frac{\lambda GM_{0}}{c_{s}^{2} D}  
\exp \left[ -\frac{t}{\tau }-\frac{\mu s+D}{\left( \mu
^{2}-1\right) c_{s}\tau }\right].
\end{equation}

Putting all the results in a single form, the perturbed density $\alpha_{3}(\vecx,t)$ at $t>0$,
for both subsonic and supersonic perturbers, is
\begin{equation}
\alpha_{3}(\vecx,t) =-\frac{\lambda \xi_{3}GM_{0}\exp(-t/\tau)}{c_{s}^{2}D},
\label{eq:alpha3c}
\end{equation}
where 
\begin{equation}
    \xi_{3}=
    \begin{dcases}
      \exp\left(\frac{\mu\tilde{s}+\tilde{D}}{1-\mu^{2}}\right) &  \hskip 0.1cm {\rm in}\hskip 0.13cm 
 {\rm Region \hskip 0.13cm 2;} \\
      2\cosh \left(\frac{\tilde{D}}{\mu^{2}-1}\right) \exp \left(-\frac{\mu\tilde{s}}{\mu^{2}-1}\right)  & \hskip 0.1cm
{\rm in} \hskip 0.13cm {\rm Region \hskip 0.13cm 2' \hskip 0.13cm and}\\
        & \hskip 0.1cm {\rm if}\hskip 0.13cm \mu>1;\\
     0 & \hskip 0.1cm {\rm otherwise,}
\label{eq:xi3}
    \end{dcases}
\end{equation}
and $\tilde{s}\equiv s/(c_{s}\tau)$ and $\tilde{D}\equiv D/(c_{s}\tau)$.
Note that $\xi_{3}\geq 0$.
We see that when $\tau\rightarrow \infty$, then both $\tilde{s}$ and $\tilde{D}$ $\rightarrow 0$.
Therefore, $\xi_{3}\rightarrow 1$ in Region 2 and
$\xi_{3}\rightarrow 2$ in Region 2$^{\prime}$, and the constant-mass case is recovered.

In contrast to $\xi_{1}$ and $\xi_{2}$, which are bound quantities ($\xi_{1}$ and $\xi_{2}$ are
less than or equal to $2$), $\xi_{3}$ is not bound. However, algebraic calculations show that
the factor $\xi_{3}\exp(-t/\tau)$, which appears in Equation (\ref{eq:alpha3c}), can only
take values between $0$ and $1$ in 
Region 2,  and between $0$ and $2$ in Region 2$^{\prime}$.
In the following, we outline how to find the extreme values of the function $\xi_{3}$ in the
different regions.

Suppose that $\mu<1$. The maxima of $\xi_{3}$ occur at locations where $\mu s+D$ are also maxima.
It is easy to show that $\mu s+D$ reaches its maximum value, $(1-\mu^{2}) c_{s}t$,  
along the sonic surface. Thus, we get from Equation (\ref{eq:xi3}) that $\xi_{3}\leq \exp(t/\tau)$ 
if $\mu<1$.

Now, consider the supersonic case ($\mu>1$). In Region 2, 
$\xi_{3}$ attains local maxima at the points where $\mu s+D$ attains local minima. The 
absolute minimum value of $\mu s+D$ in Region 2 is $-(\mu^{2}-1)c_{s}t$, and this occurs 
in all the points located on the sonic sphere satisfying $z\leq c_{s}t/\mu$. Therefore, 
$\xi_{3}\leq \exp(t/\tau)$ in Region 2 if $\mu>1$.

Finally, we consider the absolute maximum value acquired by $\xi_{3}$ within the Mach cone
(Region 2$^{\prime}$) for supersonic perturbers. 
Simple algebraic manipulations show that the absolute maximum of $\xi_{3}$ occurs at the point 
with coordinates $z=c_{s}t/\mu$ and $R=(\mu^{2}-1)^{1/2}c_{s}t/\mu$. At that position, we have
that $\xi_{3}=2\exp(t/\tau)$. 
In summary, $0\leq \xi_{3}\exp(-t/\tau)\leq 1$ in Region 2 and 
$0\leq \xi_{3}\exp(-t/\tau)\leq 2$ in Region 2$^{\prime}$.

\section{Comparison with hydrodynamical simulations}
\label{sec:appC}
\begin{figure}
\centering
 \includegraphics[width=0.5\textwidth,height=6.5cm]{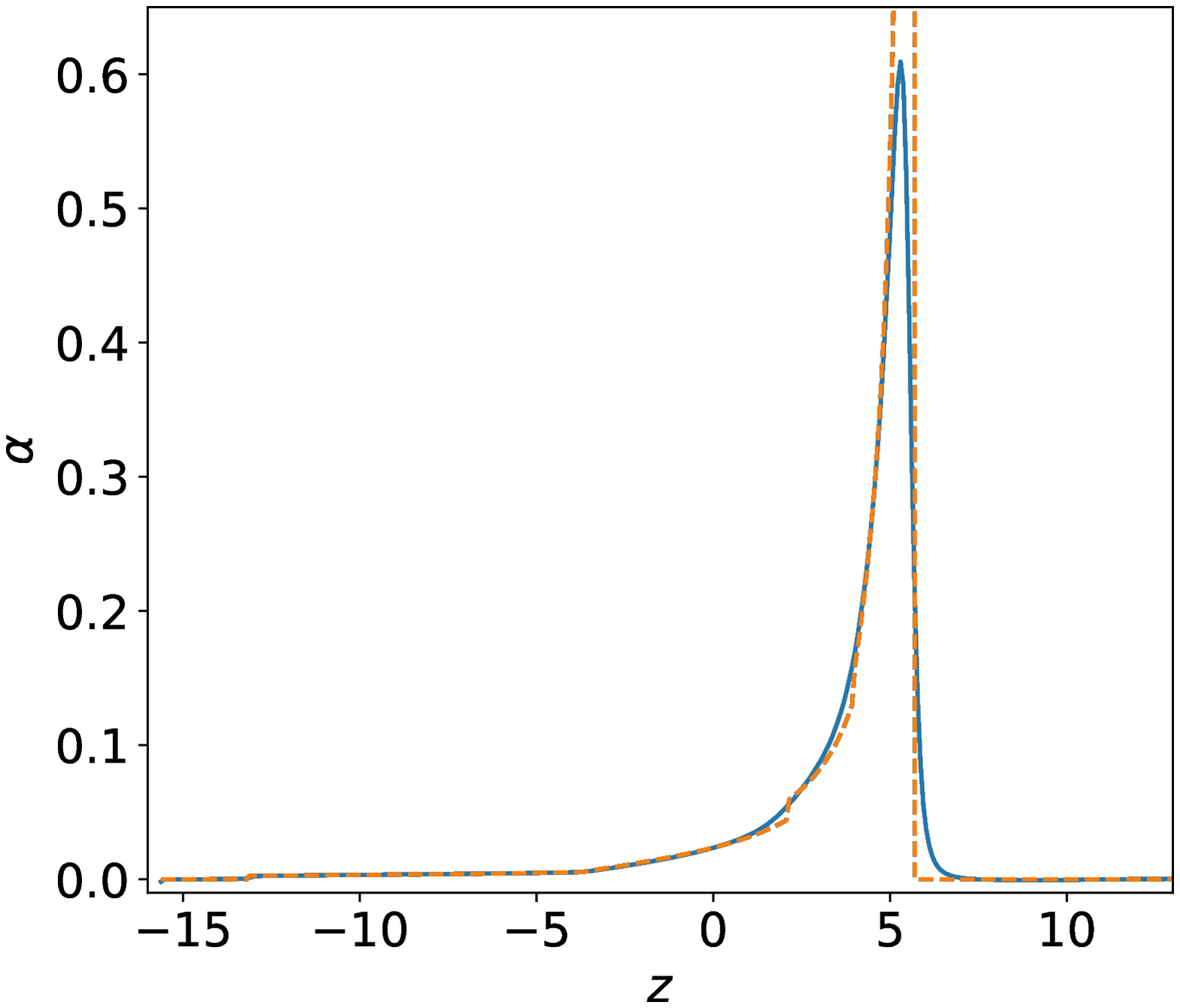}
 \caption{Comparison of the analytical results in linear theory (dashed line)
to hydrodynamical simulations (solid line) for the perturbed density $\alpha$
along the cylindrical axis $R=0$, at $t=t_{0}$. We have used the following parameters
$\mu=1.5$, $\lambda=3$, $\chi=16.1$ and $\tau=t_{0}/2$ (see Appendix \ref{sec:appC}
for details). 
}     
 \label{fig:cut_lambda3}
\end{figure}

In order to verify that the analytical derivation of the perturbed density 
described in the previous Appendices is correct, we have carried out full 
hydrodynamical simulations using the ZEUS code \citep{sto92}. A good match
between the analytical expressions and simulations was found. For illustration,
we present here the results of the wake created by
a varying-mass Plummer perturber moving at $\mu=1.5$ in an adiabatic simulation. 
In this numerical model, we have used the following parameters in code units: 
$GM_{0}/c_{s}^{2}=0.05$, $R_{\rm soft}=0.35$, $\lambda=3$, $\chi=16.1$ and $\tau=t_{0}/2$.
Our grid model has a resolution of $4.5$ zones per $R_{\rm soft}$, but we tested convergence
of the results by doubling the number of grid zones per direction.
Figure \ref{fig:cut_lambda3} shows $\alpha$ along a cut at $R=0$ (i.e. along
the $z$-axis) at $t=t_{0}$.
The deviations between analytical results and numerical ones are due to the fact 
that the perturber in our simulations is 
modelled as a Plummer sphere rather than a point mass. 

\section{Perturbed density in a more general case}
\label{sec:appD}
It is simple to extend the derivation of $\alpha$
for a more general form of the evolution of the mass of the perturber.
Suppose that $\eta (t)$ can be written as
\begin{equation}
\eta (t)= \Theta (t+t_{0}) +  {\mathcal{G}}(t) \Theta (t),
\label{eq:etaplus_gen}
\end{equation}
where ${\mathcal{G}}(t)$ is a certain function of time to
be specified. Following Appendices \ref{sec:appA} and \ref{sec:appB}, the wake induced by
the perturber is given by $\alpha=\alpha_{1}+\alpha_{\scriptscriptstyle \mathcal{G}}$, where $\alpha_{1}$ is
given in Equation (\ref{eq:alpha1appA}) and
\begin{equation}
    \alpha_{\scriptscriptstyle \mathcal{G}}=
    \begin{dcases}
     \frac{GM_{0}}{c_{2}^{2}D} \mathcal{G}\left(t+\frac{\mu s+D}{(\mu^{2}-1) c_{s}}\right) &  \hskip 0.1cm {\rm in}\hskip 0.13cm 
 {\rm Region \hskip 0.13cm 2;} \\
      \frac{GM_{0}}{c_{2}^{2}D} \sum_{j=\{-1,1\}} \mathcal{G}\left(t+\frac{\mu s+jD}{(\mu^{2}-1) c_{s}}\right)  
& \hskip 0.1cm
{\rm in} \hskip 0.13cm {\rm Region \hskip 0.13cm 2' \hskip 0.13cm and}\\
        & \hskip 0.1cm {\rm if}\hskip 0.13cm \mu>1;\\
     0 & \hskip 0.1cm {\rm otherwise.}
\label{eq:xi3}
    \end{dcases}
\end{equation}

In particular, for a perturber that loses mass at a constat rate $\Gamma$ beyond $t=0$, 
we have ${\mathcal{G}}=-\Gamma t$, and therefore
\begin{equation}
    \alpha_{\scriptscriptstyle \mathcal{G}}=
    \begin{dcases}
     -\frac{GM_{0}\Gamma}{c_{2}^{2}D} \left(t+\frac{\mu s+D}{(\mu^{2}-1) c_{s}}\right) &  
\hskip 0.1cm {\rm in}\hskip 0.13cm 
 {\rm Region \hskip 0.13cm 2;} \\
      -\frac{2GM_{0}\Gamma}{c_{2}^{2}D} \left(t+\frac{\mu s}{(\mu^{2}-1) c_{s}}\right)  
& \hskip 0.1cm
{\rm in} \hskip 0.13cm {\rm Region \hskip 0.13cm 2' \hskip 0.13cm and}\\
        & \hskip 0.1cm {\rm if}\hskip 0.13cm \mu>1;\\
     0 & \hskip 0.1cm {\rm otherwise.}
\label{eq:xi3}
    \end{dcases}
\end{equation}

\end{document}